\documentclass[journal=jctcce]{achemso}

\usepackage{amsmath,bm}
\usepackage{amssymb}
\usepackage{mathrsfs}
\usepackage{amsfonts}
\usepackage{graphicx}
\usepackage{siunitx}
\usepackage[version=3]{mhchem}
\usepackage[acronym]{glossaries}

\DeclareSIUnit\angstrom{\text {Å}}
\DeclareSIUnit\debye{\text {D}}
\DeclareSIUnit\au{\text {au}}

\newacronym{cboa}{CBOA}{cavity Born-Oppenheimer approximation}
\newacronym{bo}{BOA}{Born-Oppenheimer approximation}
\newacronym{cbohf}{CBO-HF}{cavity Born-Oppenheimer Hartree-Fock}
\newacronym{scf}{SCF}{self-consistent field}
\newacronym{dse}{DSE}{dipole self-energy}
\newacronym{pes}{PES}{potential energy surface}
\newacronym{cpes}{cPES}{cavity potential energy surface}
\newacronym{vsc}{VSC}{vibrational-strong coupling}
\newacronym{esc}{ESC}{electronic-strong coupling}
\newacronym{lp}{LP}{lower polariton}
\newacronym{up}{UP}{upper polariton}
\newacronym{ejcm}{EJCM}{extended molecular Jaynes-Cummings model}

\title{Ab-Initio Vibro-Polaritonic Spectra in Strongly Coupled Cavity-Molecule Systems}

\author{Thomas Schnappinger}
\email{thomas.schnappinger@fysik.su.se}
\affiliation{Department of Physics, Stockholm University, AlbaNova University Center, SE-106 91 Stockholm, Sweden}

\author{Markus Kowalewski}
\email{markus.kowalewski@fysik.su.se}
\affiliation{Department of Physics, Stockholm University, AlbaNova University Center, SE-106 91 Stockholm, Sweden}
\date{\today}%

\begin{document}

\begin{abstract}
Recent experiments have revealed the profound effect of strong light-matter interactions in optical cavities on the electronic ground state of molecular systems. This phenomenon, known as vibrational strong coupling (VSC), can modify reaction rates and induce the formation of molecular vibrational polaritons, hybrid states involving both photon modes and vibrational modes of molecules. We present an \textit{ab-initio} methodology, based on the cavity Born-Oppenheimer Hartree-Fock ansatz, which is specifically powerful for ensembles of molecules, to calculate vibro-polaritonic IR spectra. This method allows a comprehensive analysis of these hybrid states. Our semi-classical approach, validated against full quantum simulations, reproduces key features of the vibro-polaritonic spectra. The underlying analytic gradients also pave the way for optimizing cavity-coupled molecular systems and performing semi-classical dynamics simulations.
\end{abstract}

\maketitle
\section{Introduction}

Strong light-matter interactions between a molecular system and the electromagnetic field of an optical cavity offer new possibilities to modify chemical reactivity and selectivity, as demonstrated in recent experiments~\cite{Hutchison2012-od,George2016-sy,Thomas2016-fy,Thomas2019-ve,Hirai2020-pa,Hirai2020-uv,Kumar2023-pv,Ahn2023-qk,Bhuyan2023-se,Zhong2023-lq,Gu2023-uq,Weight2023-ma}. A particularly intriguing but not yet fully understood situation occurs when a cavity mode is strongly coupled to molecular vibrations, called \gls{vsc}~\cite{Nagarajan2021-sl,Dunkelberger2022-oh,Simpkins2023-ze,Campos-Gonzalez-Angulo2022-gb,Sidler2022-cg,Sidler2023-vm,Schnappinger2023-hh}. For \gls{vsc} the rate constants of the ground state reactions can be modified even without external driving, i.e. without explicitly adding photons into the cavity. Probably one of the most striking features observed in such experiments is the change of the vibrational spectra due to the formation of molecular vibrational polaritons, hybrid states that involve both cavity modes and vibrational modes of molecules. These vibrational polaritons are characterized by the appearance of a \gls{lp} transition and a \gls{up} transition separated by the Rabi splitting frequency $\Omega_{R}$ originating from a single vibrational peak in the spectrum of the coupled molecular-cavity system. Both the presence of vibrational polaritons and the change in chemical reactivity are characterized by a sharp resonance~\cite{li2021cavity,schafer2021shining,lindoy2023quantum}, which can be achieved if one of the cavity frequencies $\omega_{c}$ is in resonance with a vibrational mode in the reaction mixture.

Despite the large number of theoretical studies in the literature, the understanding of the underlying microscopic and macroscopic physical mechanisms, especially with respect to the effects of \gls{vsc}, is still incomplete and under discussion~\cite{Schutz2020-en,Simpkins2023-ze,Campos-Gonzalez-Angulo2022-gb,Sidler2022-cg,Sidler2023-vm,Schnappinger2023-hh,Davidsson2023-yp,Davidsson2020-bs}. To better understand the formation of molecular vibrational polaritons and their role in polaritonic chemistry, a reliable theoretical description is needed that can accurately treat also the in practice most relevant case of collective strong coupling. One way to achieve such a description is to simulate molecules in the \gls{vsc} regime in the time domain, capturing the dynamics of the system to analyze optical spectra and chemical reactivity~\cite{Flick2018-ns,Schafer2022-id,Yu2022-px,Mondal2023-ho}. These explicit time domain calculations have their advantages in simulating complex and anharmonic dynamics, but suffer from high computational cost, especially when the dynamics is calculated in a \textit{ab-initio} framework. Following the idea of Bonini and Flick~\cite{Bonini2022-ee}, the use of the generalized force-constant matrix, often called the mass-weighted Hessian matrix $\bm{H}^{M}$ in quantum chemistry, offers an alternative way to obtain vibro-polaritonic IR spectra without simulating the temporal evolution of the system. Within the well-known harmonic approximation, the eigenvectors and eigenvalues of $\bm{H}^{M}$ give the vibro-polaritonic normal modes of the correlated matter-photon system and the frequencies of the vibro-polaritons. 

In our recent work~\cite{Schnappinger2023-hh} we have introduced a Hartree-Fock ansatz in the \gls{cboa}~\cite{flick2017atoms,flick2017cavity,Flick2018-ns,Fischer2023-pe}, capable of describing the electronic ground state of single molecules as as well as of an ensemble of molecules coupled to an optical cavity. Within the framework of this \gls{cbohf} ansatz, we now derive analytic expressions for the first derivatives of the energy with respect to the nuclear and photonic degrees of freedom. These analytic gradients are then used to construct the mass-weighted Hessian matrix $\bm{H}^{M}$ via finite differences and subsequently \textit{ab-initio} vibro-polaritonic IR spectra within the harmonic approximation are calculated. After introducing the \gls{cbohf} formalism, the first part of this work describes how the analytic expressions for the first derivatives of the energy are obtained and used to calculate vibro-polaritonic IR spectra in the \gls{cbohf} framework. Next, the vibro-polaritonic IR spectra in the harmonic approximation are carefully compared against a full quantum mechanical approach for the rather anharmonic case of one and two diatomic hydrogen fluoride \ce{HF} molecules. Even in this rather simple molecular example, we observe that the interaction between an optical cavity and the molecule(s) leads to a detuning and a change in the optical length of the cavity for both the harmonic approximation and the full quantum treatment, in agreement with our recent findings~\cite{Sidler2023-vm}. Since both the \gls{scf} treatment and the correct description of the full \gls{dse} contribution are crucial to capture relevant aspects for the description of strongly coupled molecules and their chemical properties~\cite{Schafer2020-cb,Sidler2022-cg,Sidler2023-vm,Schnappinger2023-hh,Ruggenthaler2023-aa}, we analyze their direct influence on the vibro-polaritonic IR spectra of small ensembles of \ce{HF} molecules. In the last part, we leave the case of diatomic molecules and discuss the vibro-polaritonic IR spectra for ammonia (\ce{NH3}). Here, the interaction with the confined light mode of an optical cavity influences the whole vibrational spectrum not only by creating vibrational polaritons but also by allowing symmetry breaking.

\section{Theory}

To describe the interaction of a confined light mode with atoms, molecules, and condensed matter systems, the non-relativistic Pauli–Fierz Hamiltonian~\cite{spohn2004dynamics,Ruggenthaler2018-ew,jestadt2019light,lindoy2023quantum,Ruggenthaler2023-aa} in the length gauge, assuming the dipole approximation is an accurate starting point. If the quantized cavity modes are coupled via their characteristic frequencies to vibrational degrees of freedom of molecules, the situation is described as \gls{vsc}, for which the \gls{cboa}~\cite{flick2017atoms,flick2017cavity,Flick2018-ns,Fischer2023-pe} is a well suited theoretical approach. 
Within the \gls{cboa}, the cavity modes are grouped with the nuclei in a generalized Born-Huang expansion~\cite{Schafer2018-vf,Ruggenthaler2023-aa}, and then one can subsequently solve the quantum problem of the electrons and then of the nuclei and photons. Here we first focus on the electronic degrees of freedom that have been shown to be strongly affected even for cavity modes that are tuned to the vibrational degrees of freedom~\cite{Sidler2023-vm,Schnappinger2023-hh}.
In the following bold symbols denote vectors, and atomic units ($\hbar=4\pi; \varepsilon_0=m_e=1$) are used throughout the paper, unless otherwise noted. 
For a single mode cavity the electronic \gls{cboa} Hamiltonian takes the form
\begin{equation}
\label{eq:h_cbo}
\begin{split}
\hat{H}_{CBO} &= \hat{H}_{el} + \frac{1}{2} \omega_c^2 q_{c}^2 - \omega_c q_{c} \left(\bm{\lambda}_{c} \cdot \bm{\hat{\mu}} \right) + \frac{1}{2} \left(\bm{\lambda}_{c} \cdot \bm{\hat{\mu}} \right)^2\,, 
\end{split}
\end{equation}
where 
\begin{equation}
\begin{split}
\quad \bm{\hat{\mu}} = \bm{\hat{\mu}}_{el} + \bm{\mu}_{Nuc} = -\sum_{i=1}^{N_{el}} \bm{\hat{r}}_i  + \sum_{A=1}^{N_{Nuc}} Z_{A} \bm{R}_A\,,
\end{split}
\end{equation}
represents the molecular dipole operator, which is defined by the operators of the $N_{el}$ electron coordinates $\bm{\hat{r}}$, the classic coordinates $\bm{R}$ of the $N_{Nuc}$ nuclei and the nuclear charges $Z_A$.
$\hat{H}_{el}$ is the Hamiltonian for the field-free many-electron system, and the second term defines the harmonic potential introduced by the photon displacement field, with the photon displacement coordinate $q_{c}$ and $\omega_c $ being the frequency of the cavity mode. 
The third term of Eq.~\eqref{eq:h_cbo} describes the dipole coupling between the molecular system and the photon displacement field, which is characterized by the coupling strength $\bm{\lambda}_{c}$. 
The last term is the \gls{dse} operator~\cite{Rokaj2018-ww,Schafer2020-cb,Sidler2022-cg}, which is an energy contribution that describes the self-polarization of the molecule-cavity system. 
The coupling parameter $\bm{\lambda}_{c}$ for a cavity with an effective mode volume $V_{c}$ is defined as follows:
 \begin{equation}
 \label{eq:lam}
\bm{\lambda}_{c} =  \bm{e} \lambda_{c} =  \bm{e}  \sqrt{\frac{4 \pi}{V_{c}}}\,.
\end{equation}
The unit vector $\bm{e}$ denotes the polarization axis of the cavity mode. 

\subsection{CBO-Hartree-Fock}
In our recent work~\cite{Schnappinger2023-hh} we have introduced the \gls{cbohf} approach representing a formulation of the well-known Hartree-Fock ansatz in the context of the \gls{cboa}. The resulting energy expectation value $\bigl\langle  E_{CBO} \bigr\rangle$ consists of four energy contributions:
\begin{equation}
\label{eq:e_cbo_part}
\bigl\langle  E_{CBO} \bigr\rangle  = E_{el} +  E_{lin} + E_{dse} + E_{dis} \quad  \text{ with} \quad E_{dis} = \frac{1}{2} \omega_c^2 q_{c}^2\,.
\end{equation}
The first term $E_{el}$ contains all Hartree-Fock energy components of the many-electron system~\cite{aszabo82-qc,Jensen2017-fq} and $E_{lin}$ describes the linear dipole coupling between the photon displacement field, the electrons, and the nuclei. The energy contribution $E_{dse}$ is due to the \gls{dse} operator in the Hamiltonian and $E_{dis}$ is the energy resulting from the photon displacement field~\cite{Schafer2020-cb}. Following the standard procedure the transformation of  $\bigl\langle  E_{CBO} \bigr\rangle$  in the basis of atomic orbitals results in 
\begin{equation}
\bigl\langle  E_{CBO} \bigr\rangle  =  \sum_{\alpha,\beta}^{M}  D_{\alpha,\beta} F_{\alpha,\beta} + \tilde{E}_{n,c} = \sum_{\alpha,\beta}^{M}  D_{\alpha,\beta}  \bigl\langle \alpha \big| \tilde{h}\big| \beta \bigr\rangle  + 
\frac{1}{2} \sum_{\alpha,\beta,\gamma,\delta}^{M} D_{\alpha,\beta} D_{\gamma,\delta} \bigl\langle \alpha \beta \big| \tilde{g} \big| \gamma \delta \bigr\rangle  +  \tilde{E}_{n,c}
\end{equation}
where $\alpha,\beta,\gamma,\delta$ denote atomic orbitals, $F_{\alpha,\beta}$ the \gls{cbohf} Fock matrix and $D_{\alpha,\beta}$ the corresponding density matrix elements. Due to the classical nature of the nuclei and the photon displacement field in the \gls{cboa}, when we determine the electronic ground state, all their energy contributions are scalar and summed up in $\tilde{E}_{n,c}$:
\begin{equation}
\tilde{E}_{nuc,c}  = V_{nn} + E_{dis} +  E^{(nuc)}_{lin} + E^{(nuc)}_{dse}.
\end{equation}
The modified one-electron integrals $\bigl\langle \alpha \big| \tilde{h}\big| \beta \bigr\rangle$ and two-electron integrals $\bigl\langle \alpha \beta \big| \tilde{g} \big| \gamma \delta \bigr\rangle$ used to build the \gls{cbohf} Fock matrix elements $F_{\alpha,\beta}$ consist of the standard one-electron integral $\bigl\langle \alpha \big| h \big| \beta \bigr\rangle$ and two-electron integrals $\bigl\langle \alpha \beta \big| g \big| \gamma \delta \bigr\rangle$ as well as terms describing the linear cavity-electron interaction and the electronic \gls{dse} contributions:
\begin{equation}
\label{eq:1e_2e_ao}
\begin{split}
\bigl\langle \alpha \big| \tilde{h}\big| \beta \bigr\rangle & = \bigl\langle \alpha \big| \hat{h} \big| \beta \bigr\rangle  - \omega_c q_{c}
\bigl\langle \alpha \big| \bm{\lambda}_{c} \cdot \bm{\hat{r}} \big| \beta \bigr\rangle -  
\bm{\lambda}_{c} \cdot \bm{\mu}_{Nuc}  \bigl\langle \alpha \big| \bm{\lambda}_{c} \cdot \bm{\hat{r}} \big| \beta \bigr\rangle
+ \frac{1}{2} \bigl\langle \alpha \big|  \left( \bm{\lambda}_{c} \cdot \bm{\hat{r}} \right)^2 \big| \beta \bigr\rangle \\ 
 \bigl\langle \alpha \beta \big| \tilde{g} \big| \gamma \delta \bigr\rangle & = \bigl\langle \alpha \beta \big| g \big| \gamma \delta  \bigr\rangle + \bigl\langle \alpha \big| \bm{\lambda}_{c} \cdot \bm{\hat{r}} \big| \beta \bigr\rangle  \bigl\langle \gamma \big|  \bm{\lambda}_{c} \cdot \bm{\hat{r}} \big| \delta \bigr\rangle - \bigl\langle \alpha \big| \bm{\lambda}_{c} \cdot \bm{\hat{r}} \big| \delta \bigr\rangle \bigl\langle \gamma \big| \bm{\lambda}_{c} \cdot \bm{\hat{r}} \big| \beta \bigr\rangle 
\end{split}
\end{equation}

\subsection{CBO-Hartree-Fock Gradients}
The first derivative of the energy $\bigl\langle  E_{CBO} \bigr\rangle$ with respect to a nuclear or photon displacement coordinate $\zeta_i$ is
\begin{equation}
\label{eq:def_grad}
 \frac{\partial}{\partial \zeta_i} \bigl\langle  E_{CBO} \bigr\rangle = \sum_{\alpha,\beta}^{M}  D_{\alpha,\beta}  \frac{\partial}{\partial \zeta_i} F_{\alpha,\beta} + \sum_{\alpha,\beta}^{M} F_{\alpha,\beta}  \frac{\partial}{\partial \zeta_i} D_{\alpha,\beta}   + \frac{\partial}{\partial \zeta_i} \tilde{E}_{n,c}.
\end{equation}
The first term in Eq.~\eqref{eq:def_grad} is the Hellmann-Feynman term, the second term is the wavefunction derivative, or Pulay term, and the last part is the derivative of all scalar energy contributions. As the \gls{cbohf} wave function is variational optimized, the explicit calculation of the density derivatives can be avoided~\cite{Pulay1969-hg} and the Pulay term can be written in terms of orbital energies $\epsilon_i$, orbital coefficients $c_i$ and overlap integral derivatives:
\begin{equation}
\label{eq:devDensity}
\begin{split}
        \sum_{\alpha,\beta}^{M} F_{\alpha,\beta} \frac{\partial D_{\alpha,\beta} }{\partial \zeta_i}  &= - \sum_{\alpha,\beta}^{M} W_{\alpha,\beta}  \frac{\partial S_{\alpha,\beta} }{\partial \zeta_i}\\
        &= \sum_{\alpha,\beta}^{M}  \sum_{i=1}^{N_{oc}} \epsilon_i c_{i,\alpha}^{*}c_{i,\beta} \left( 
\bigl\langle \frac{\partial \alpha}{\zeta_i}  \big| \beta \bigr\rangle +\bigl\langle \alpha \big| \frac{\partial \beta}{\zeta_i} \bigr\rangle  \right)
\end{split}
\end{equation}
In the case of $\zeta_i$ being a photon displacement coordinate, Eq.~\eqref{eq:devDensity} becomes zero, since the atomic orbitals $\alpha$, $\beta$ are independent of $q_c$. A detailed derivation of all "new" terms in Eq.~\eqref{eq:def_grad} for $\zeta_I = R_{i}$ and $\zeta_I = q_c$ can be found in section~S1 of the supporting informations. All derivatives for $\bigl\langle\hat{h}\bigl\rangle$, $\bigl\langle\hat{g}\bigl\rangle$, and $V_{nn}$ with respect to nuclear coordinates are well known and can be found elsewhere in the literature, for example~\cite{Jensen2017-fq}. The resulting analytic \gls{cbohf} gradient $\bm{g}_{CBO}$ is a $\left( 3N_{A} +1 \right)$ vector with $N_{A}$ being the number of atoms in the molecule:
\begin{equation}
\bm{g}_{CBO} =   \begin{pmatrix} 
    \frac{\partial}{\partial R_{1}}  & \frac{\partial}{\partial R_{2}} & \hdots & \frac{\partial}{\partial q_c} 
 \end{pmatrix}^{T} \bigl\langle  E_{CBO} \bigr\rangle
\end{equation}
 
\subsection{CBO-Hartree-Fock Frequencies}
In order to calculate frequencies in the harmonic approximation for the \gls{cbohf} ansatz we need to calculate the Hessian matrix, $\bm{H}$, with dimensions $\left(3N_{A} +1\right)\text{x}\left(3N_{A} +1\right)$ for the case of a single photon mode. Each element $\bm{H}_{ij}$ is the second derivative of $\bigl\langle  E_{CBO} \bigr\rangle$ with respect to nuclear coordinates respectively the photon displacement coordinate~\cite{Bonini2022-ee} that is
\begin{equation}
\bm{H}_{ij} = \bm{H}_{ji} = \frac{\partial^2}{\partial \zeta_{i} \partial \zeta_{j}} \bigl\langle  E_{CBO} \bigr\rangle.
\end{equation}
Using the analytic \gls{cbohf} gradient $\bm{g}_{CBO}$ the second derivatives can be determined using finite differences:
\begin{equation}
\bm{H}_{ij} = \frac{\partial}{\partial \zeta_{i}} \bm{g}_{cbo}^j \approx \frac{\bm{g}_{cbo}^j\left(\zeta_{i}+\Delta \right)-\bm{g}_{cbo}^j\left(\zeta_{i}-\Delta \right)}{2\Delta}
\end{equation}
The harmonic frequencies and corresponding normal modes can be obtained by transforming the Hessian matrix $\bm{H}$ from the $\zeta_i$ coordinates into their mass-weighted version $\bm{H}^{M}$ and subsequently solving the eigenvalue problem:
\begin{equation}
\bm{A}^{\dagger}\bm{H}^{M}\bm{A} = \bm{\mathcal{L}}
\end{equation}
Here $\bm{\mathcal{L}}$ is the diagonal matrix of eigenvalues $\ell_i$ and $\bm{A}$ is the matrix that diagonalizes the mass-weighted Hessian $\bm{H}^{M}$. $\bm{A}$ is formed by the juxtaposition of its eigenvectors $\bm{a}_i$ which define the normal modes. From the $\left( 3N_{A} +1 \right)$ eigenvalues, only $\left( 3N_{A} +1 -6 \right)$, or $\left( 3N_{A} +1 -5\right)$ if the molecule is linear, correspond to the harmonic frequencies of vibrational motions, whereas the others correspond to translations and rotations. The harmonic frequencies, in \unit{\per\centi\meter}, are given by:
\begin{equation}
^{H}\nu_i = \sqrt{ \frac{\ell_i}{ 4 \pi^2 c^2} }
\end{equation}
The corresponding intensities $^{H}\mathcal{I}_i$ in the harmonic approximation are calculated as the projection of the dipole moment gradient on the normal mode vectors $\bm{a}_i$:
\begin{equation}
^{H}\mathcal{I}_i  = \left( \bm{\nabla} \bigl\langle \bm{\hat{\mu}} \bigr\rangle \cdot \bm{a}_i \right)^2
\end{equation}
The necessary dipole moment gradient $\bm{\nabla} \bigl\langle \bm{\hat{\mu}} \bigr\rangle$ is calculated via finite differences. To validate the quality of the harmonic approximation and to go beyond the semi-classical treatment we determined the vibartional spectra of the cavity coupled systems by then also determining the quantum mechanical wavefunction of the nuclei and photons in the Born-Huang expansion. Therefore, we not only have access to the fundamental transitions but also to overtones. The anharmonic frequency $^{A}\nu_{j}$ for a given fundamental transition is the difference of the energies of the eigenstate $j$ and the nuclear-photonic ground state. The corresponding intensity $^{A}\mathcal{I}_{j}$ are calculated with the nuclear-photonic eigenfunctions $\chi_i$ of the coupled molecular-cavity system:
\begin{equation}
^{A}\mathcal{I}_{0j} = \bigl\langle \chi_0 \big|\bm{\hat{\mu}}\big| \chi_j \bigr\rangle
\end{equation}

\section{Computational Details}

The analytic gradients for the \gls{cbohf} ansatz have been implemented in the Psi4NumPy environment~\cite{Smith2018-tu}, which is an extension of the PSI4~\cite{Smith2020-kq} electronic structure package. 
All calculations were performed using the aug-cc-pVDZ basis set~\cite{Kendall1992-wu} and the geometry of the isolated single \ce{HF} and \ce{NH3} molecule have been optimized at the Hartree-Fock level of theory. Consistent with our previous studies~\cite{Sidler2023-vm,Schnappinger2023-hh}, the optimized structure of a single \ce{HF} molecule is used to define small ensembles of up to four molecules to study the collective effects on the infrared spectra of molecules in cavities. These ensembles are constructed by placing replicas of the single \ce{HF} molecule separated by \SI{800}{\angstrom} inside the cavity. Two different orientations of the molecular \ce{HF} ensembles are studied: The \textit{all-parallel} configuration, in which all \ce{HF} molecules are aligned parallel and the \textit{antiparallel} configuration, which is characterized by a pairwise antiparallel orientation of the $N_{mol}$ \ce{HF} molecules both with respect to the cavity polarization axis of the cavity. For \ce{NH3} only the single molecule case is studied. As discussed in the supporting information of Ref.~\citenum{Schnappinger2023-hh}, the aligned orientation of \ce{HF} is not the most energetically favorable when coupled to a cavity, but a transition state along the rotation with respect to the polarization axis. However, the two corresponding minima lead to a situation where the molecule is not coupled to the cavity field. For the case of \ce{NH3} the reoptimization in the presence of the cavity is mainly characterized by a rotation in the laboratory frame, resulting in an orientation with reduced coupling. Since the effects on the internal coordinates of \ce{HF} and \ce{NH3} are negligible for the coupling strengths studied, we do not re-optimize the geometries of the molecular systems in the cavity and align the molecules for maximum coupling to the cavity. In all \gls{cbohf} calculations performed in this work, we consider a single mode, lossless cavity. We keep the collective coupling strength $\bm{\lambda}_{c}$ constant by applying a scaling factor of ${1}/{\sqrt{N_{mol}}}$ to obtain a fixed Rabi splitting for different ensemble sizes and treat $\lambda_0$ as a tunable coupling parameter:
\begin{equation}
\label{eq:coupling}
\bm{\lambda}_{c} = \frac{\lambda_0}{\sqrt{N_{mol}}} \bm{e}
\end{equation} 
Here $\lambda_0$ is equivalent to $\lambda_c$ in Eq.~\eqref{eq:lam} in the single molecule case.
As a result, we increase the mode volume $V_c$ of the cavity, but by including more molecules, we keep the average density of molecules $N_{mol}/V_c$ fixed. We use an artificially increased coupling strength $\lambda_0$ in the range of 0.004 to 0.04, which corresponds to effective mode volumes, see Eq.~\eqref{eq:lam}, in the single-molecule case as large as \SI{125.27}{\nano\metre\cubed}~(for $\lambda_0=0.004$) or as small as \SI{1.25}{\nano\metre\cubed}~(for $\lambda_0= 0.04$). 

To validate the harmonic approximation, the vibrational spectra of a single \ce{HF} molecule and a pair of \ce{HF} molecules coupled to a cavity are calculated in a fully quantum mechanical way. The necessary nuclear-photonic eigenfunctions $\chi_i$ of the coupled molecular-cavity system are obtained by using the imaginary time propagation method~\cite{Kosloff1986-ow} with the Arnoldi propagation scheme~\cite{Smyth1998-dv} both implemented in our in-house quantum dynamics code (QDng). The used \gls{cbohf} \glspl{cpes} are represented by a two-dimensional or three-dimensional numerical grid with 128 points along the bond lengths $R$ and 64 points along the photon displacement coordinate $q_c$.
For all vibrational spectra shown in this work the underling signals are broadened by a Lorentzian function with the width of \SI{10}{\per\centi\meter}. All calculations were performed in a reproducible environment using the Nix package manager together with NixOS-QChem \cite{nix} (commit f5dad404) and Nixpkgs (nixpkgs, 22.11, commit 594ef126).

\section{Results and Discussion}

\subsection{Validation of the Harmonic Approximation}

\begin{figure}[htb!]
     \centering
         \includegraphics[width=0.9\textwidth]{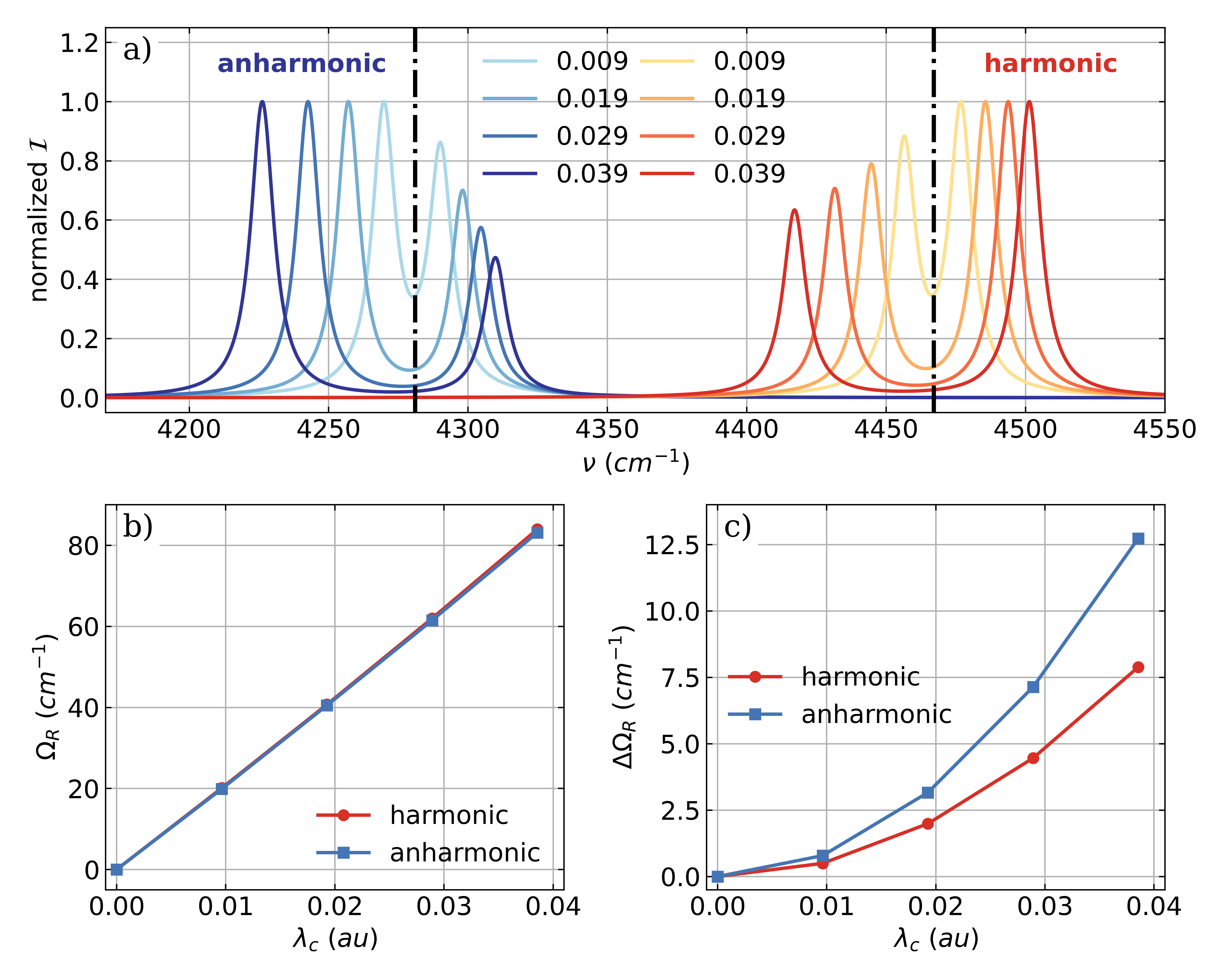}
    \caption{a) Vibro-polaritonic IR spectra of a single \ce{HF} molecule calculated in the harmonic approximation (redish) and in the full-qunatum/anharmonic setup (bluish), individually normalized for each coupling strength. Black dashed-dotted lines indicate the bare molecular frequencies of the harmonic ($^{H}\nu_{1} = \SI{4467}{\per\centi\meter}$) and anharmonic ($^{A}\nu_{1} = \SI{4281}{\per\centi\meter}$) fundamental transitions. The cavity frequency $\omega_c$ is resonant with the corresponding fundamental transition, and the coupling strength $\lambda_{c}$ is increased from \SI{0.009}{\au} to \SI{0.039}{\au} (from lightest to darkest color). The Rabi splitting $\Omega_R$ (b) and its asymmetry $\Delta \Omega_R = \omega_c - 0.5\left( \nu^{LP} + \nu^{UP}\right)$ (c) as a function of $\lambda_{c}$.} 
\label{fig:vib_pol_spec_hf}
\end{figure}

Let us start with the validation of the harmonic approximation for vibro-polaritonic IR spectra of a single \ce{HF} molecule coupled to an optical cavity. The spectra of the bare \ce{HF} molecule calculated in the harmonic approximation and fully quantum mechanical (anharmonic) are shown in Fig.~S1 in the supporting informations. The fundamental vibrational transition in the harmonic approximation has a frequency  of $^{H}\nu_{1} = \SI{4467}{\per\centi\meter}$. The fully quantum mechanical treatment leads to an expected strong red-shifted of \SI{186}{\per\centi\meter} for the fundamental vibrational transition ($^{A}\nu_{1} = \SI{4281}{\per\centi\meter}$). The vibro-polaritonic infrared spectra for different coupling strengths calculated in the harmonic approximation and full-quantum (anharmonic) as well as the trends in the Rabi splitting $\Omega_R$ are shown in Fig.~\ref{fig:vib_pol_spec_hf}. 

The resonant coupling of the cavity mode to the fundamental transition leads to the expected formation of a \gls{lp} transition and \gls{up} transition in both the harmonic approximation and full-anharmonic treatment, see Fig.~\ref{fig:vib_pol_spec_hf}~a). The same anharmonic shift as in the cavity-free spectra is observed in the vibro-polaritonic spectra due to the anharmonicity along the nuclear coordinate independent of the used coupling strength $\lambda_{c}$. In addition to the red shift, the harmonic and full-anharmonic spectra differ mainly in their intensity patterns. In the harmonic approximation, the \gls{up} transition is more intense and the difference between \gls{lp} and \gls{up} is smaller compared to the full-quantum case. For the latter, the \gls{lp} transition is more intense. This discrepancy is probably due to the fact that the intensities in the harmonic approximation are only calculated as the first derivative of the dipole moment. The increase in $\lambda_{c}$ leads to a linear increase of the Rabi splitting $\Omega_R$ between \gls{lp} and \gls{up} which is identical for both the harmonic and full-anharmonic spectra (Fig.~\ref{fig:vib_pol_spec_hf} b)). Consistent with our previous work~\cite{Sidler2023-vm}, we observe an asymmetry $\Delta \Omega_R = \omega_c - 0.5\left( \nu^{LP} + \nu^{UP}\right)$ in the Rabi splitting (Fig.~\ref{fig:vib_pol_spec_hf} c)), where the \gls{lp} is stronger red-shifted than the \gls{up} with respect to $\omega_c$. This asymmetry is observed for the harmonic approximation and the full-anharmonic treatment and in both cases $\Delta \Omega_R$ increases quadratically with $\lambda_{c}$. However, the asymmetry is more pronounced in the full-anharmonic case. This observed asymmetry of $\Omega_R$ can also be understood as an  detuning and change in the optical length of the cavity when interacting with the molecule. Similar results are found for two \ce{HF} molecules in the \textit{all-parallel} and \textit{antiparallel} configuration coupled to an optical cavity. The corresponding vibro-polaritonic spectra for both case are shown in  Fig.~S2 and Fig.~S3 of the supporting informations. Due to the rescaling of $\bm{\lambda}_{c}$ (see Eq.~\eqref{eq:coupling}) the spectra for the two molecule case look very similar to the single molecule case and both configurations are nearly indistinguishable in their spectra. 

\begin{figure}[htb!]
     \centering
         \includegraphics[width=0.8\textwidth]{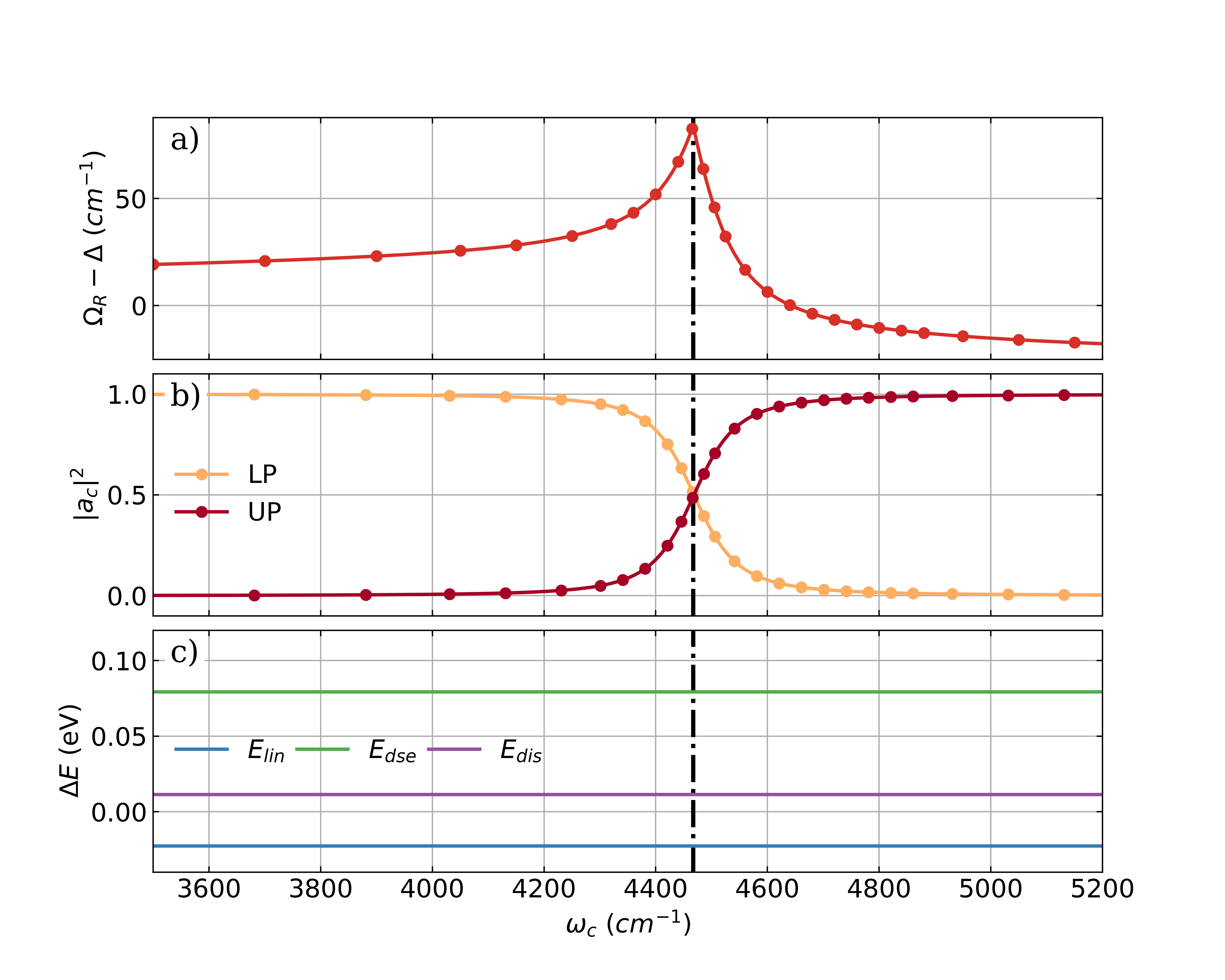}
    \caption{a) Difference between the Rabi splitting $\Omega_R$ and $\Delta =\omega_c - ^{H}\nu_{1}$ as a function of the cavity frequency $\omega_c$ for a single \ce{HF} molecule using the \gls{cbohf} harmonic approximation. Here $\Delta$ describes the difference between the cavity frequency $\omega_c$ and the fundamental bare molecular transition  $^{H}\nu_{1}$. b) The normal mode value $|a_c|^{2}$ describing the change in $q_c$ for the \gls{lp} and \gls{up} vibrational transition as function of $\omega_c$. c) Energy contributions due to the cavity interaction as function of the cavity frequency $\omega_c$ for the \gls{cbohf} ansatz. Black dashed-dotted lines indicate the frequencies of the harmonic fundamental transition ($^{H}\nu_{1} = \SI{4467}{\per\centi\meter}$). A constant coupling strength $\lambda_{c}$ of \SI{0.039}{\au} is used.} \label{fig:detuning_mixing}
\end{figure}

To further extend the validation of the harmonic approximation, we next compare the effect of detuning the cavity frequency with respect to the fundamental transition of the single \ce{HF} molecule on the vibro-polaritonic IR spectra. Besides the spectral properties, the harmonic approximation gives access to normal modes $\bm{a}_i$. In the standard \gls{bo}, these normal mode vectors describe the displacement of classical nuclei associated with the corresponding vibrational transition. In the \gls{cboa} for the case of single photon mode these vectors have an additional term $a_c$ describing the change in the classical photon displacement field $q_c$. The value of $|a_c|^{2}$ for a given normal mode is a measure of how strongly the corresponding vibrational transition interacts with the photon field. For an uncoupled light-matter system a pure molecular transition is characterized by an $|a_c|^{2}$ value of zero, while the pure photon mode has a value of one. Note that due to the length gauge $q_c$ and the corresponding value $a_c$ are no longer a pure photonic quantity if light and matter are coupled~\cite{Rokaj2018-ww,Schafer2020-cb}. Nevertheless, $|a_c|^{2}$ can still be used as a probe to identify how photonic the corresponding vibrational transition is. The obtained information is comparable with the coefficients in Hopfield models~\cite{Hopfield1958-ms}. In Fig.~\ref{fig:detuning_mixing} the Rabi splitting $\Omega_R$ in harmonic approximation, the magnitude $|a_c|^{2}$ and the cavity-induced energy changes $E_{lin}$, $E_{dse}$ and $E_{dis}$ are plotted as function of the cavity frequency $\omega_c$ for a fixed coupling strength. 

\begin{figure}[htb!]
     \centering
         \includegraphics[width=0.8\textwidth]{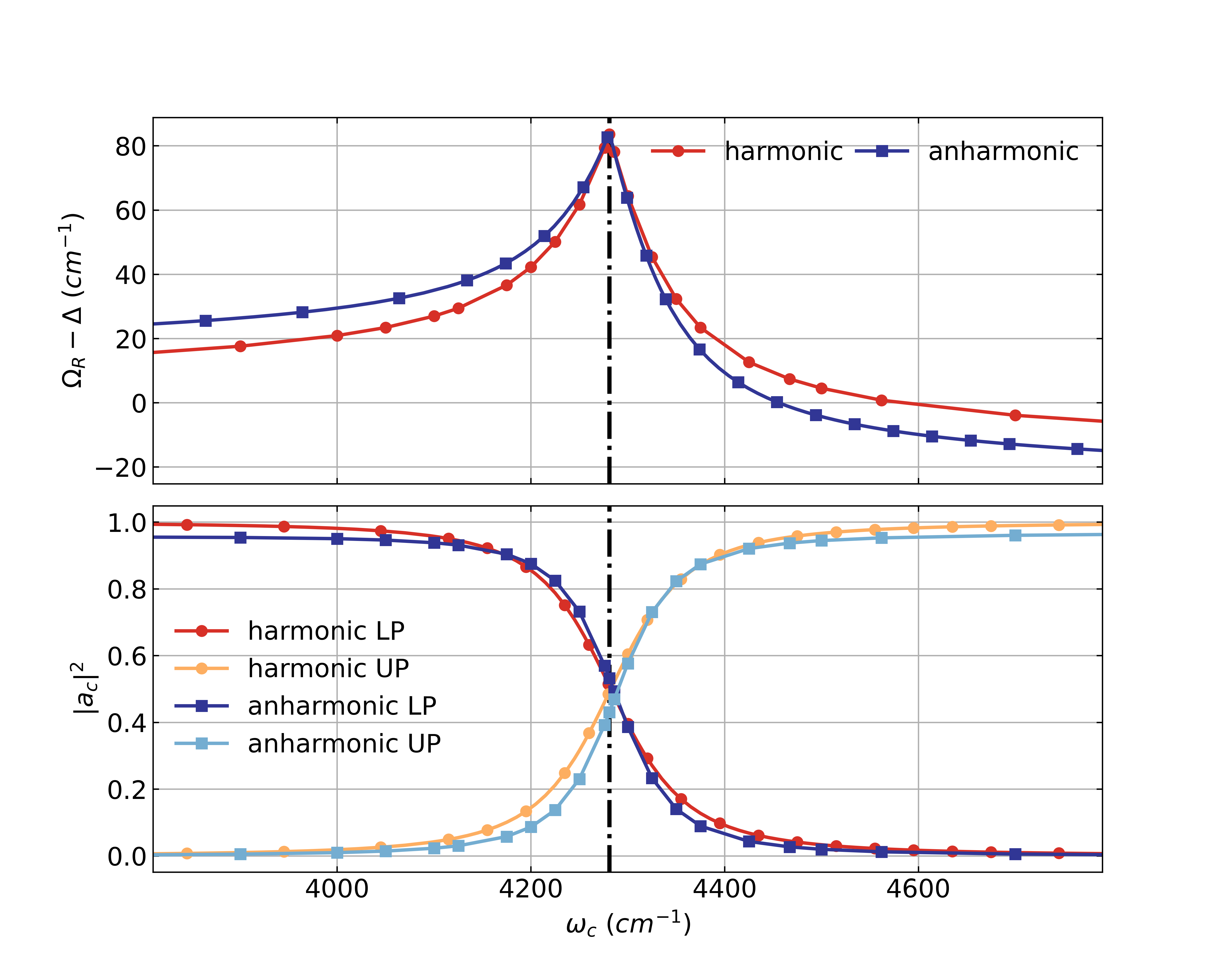}
    \caption{a) Difference between the Rabi splitting $\Omega_R$ and $\Delta =\omega_c - ^{H}\nu_{1}$ as a function of the cavity frequency $\omega_c$ for a single \ce{HF} molecule in the full anharmonic treatment (blue) and the harmonic approximation (red). Here $\Delta$ describes the difference between the cavity frequency $\omega_c$ and the fundamental bare molecular transition  $^{H}\nu_{1}$. b) Change of the cavity normal mode magnitude $|a_c|^{2}$ (red) for \gls{lp} and \gls{up} and the values $|a_{c}^{LP}|^{2}$ and $|a_{c}^{UP}|^{2}$ of the one-photon cavity vacuum state  (blue). The harmonic results are shifted to match the anharmonic fundamental transition ($^{A}\nu_{1} = \SI{4281}{\per\centi\meter}$), indicated by the black dashed dotted line. A constant coupling strength $\lambda_{c}$ of \SI{0.039}{\au} is used.} \label{fig:detuning_mixing_comparison}
\end{figure}

For better visualization, the difference between $\Omega_R$ and $\Delta =\omega_c - ^{H}\nu_{1}$, which describes the detuning of the cavity mode, is shown in Fig.~\ref{fig:detuning_mixing_comparison}~a). For an uncoupled molecular cavity system, this difference would be zero and independent of $\omega_c$. Note that $\Omega_R-\Delta$ is only a measure of the total size of $\Omega_R$, not of the asymmetry $\Delta \Omega_R$, which is discussed in Fig.~\ref{fig:vib_pol_spec_hf}. As expected, the largest Rabi splitting $\Omega_R$ is obtained for $\omega_c$ being resonant with the fundamental bare molecular transition ($^{H}\nu_{1}$). Interestingly, the difference between $\Omega_R$ and the detuning $\Delta$ is not symmetric with respect to ${H}\nu_{1}$, and tends to a finite non-zero value even for large detunings. These results clearly show that the observation of a large Rabi splitting is a rather sharp resonance effect. A similar resonance is visible for the normal mode values $|a_c|^{2}$ of the vibrational transitions of the \gls{lp} state and \gls{up} state, as shown in Fig.~\ref{fig:detuning_mixing_comparison}~b). For $\omega_c$ equal to $^{H}\nu_{1}$ the $|a_c|^{2}$ values of \gls{lp} and \gls{up} are close to $0.5$, indicating that both vibrational transitions have a hybrid light-matter character. In contrast to the prominent resonance features observed in Fig.~\ref{fig:detuning_mixing_comparison}~a) and b), the cavity-induced energy modifications $E_{lin}$, $E_{dse}$ and $E_{dis}$ shown in Fig.~\ref{fig:detuning_mixing_comparison}~c) are constant for all values of $\omega_c$. This result indicates that the energy changes induced by the cavity do not depend on any resonance condition, consistent with recent theoretical findings~\cite{Pavosevic2023-lf,Pavosevic2023-vc}, where the coupling to a non-resonant cavity alters the ground state \gls{pes}. Therefore, the experimentally observed change in chemical reactivity near resonance conditions cannot be explained by simple shifts of the underlying potential energy surfaces alone. 

The direct comparison of the Rabi splitting and the light-matter character of the \gls{lp} and \gls{up} transitions between the harmonic and full anharmonic simulations is shown in Fig.~\ref{fig:detuning_mixing_comparison}. Since we treat the coupled cavity-molecular system in full quantum fashion, we have access to the nuclear-photonic eigenfunctions of the \gls{lp} state ($\chi_{LP}$) and the \gls{up} state ($\chi_{UP}$). Figs.~S5 and~S6 showing examples of uncoupled and coupled nuclear-photonic eigenfunctions can be found in section~S3 of the supporting information. The coupled eigenfunctions can be expanded in terms of the uncoupled pure matter eigenfunction ($\chi_{M}^{(0)}$) and the photonic eigenfunction ($\chi_{C}^{(0)}$):
\begin{equation}
    \begin{split}
\chi_{LP} & = a_{M}^{LP} \chi_{M}^{(0)} + a_{c}^{LP} \chi_{c}^{(0)} \\
\chi_{UP} & = a_{M}^{UP} \chi_{M}^{(0)} + a_{c}^{UP} \chi_{c}^{(0)}
\end{split}
\end{equation}
The absolute square of obtained expansion coefficients $|a_{c}^{LP}|^{2}$ and $|a_{c}^{UP}|^{2}$ can be used as a measure of the light-matter character similar to their classical pendant normal mode values $|a_c|^{2}$. For a better comparison, the results of the harmonic simulations are red-shifted by \SI{186.0}{\per\centi\meter} and the black dashed dotted line in Fig.~\ref{fig:detuning_mixing_comparison} indicates the anharmonic fundamental transition.

In good agreement with the previous discussion, the results obtained using the harmonic approximation are very similar to the results of the full-anharmonic simulations. After accounting for anharmonicity, the differences between $\Omega_R$ and detuning $\Delta$ (Fig.~\ref{fig:detuning_mixing_comparison}~a)) are close to resonance nearly identical and for large detunings qualitatively similar. The asymmetry $\Delta \Omega_R$, as well as the asymmetry of the difference between $\Omega_R$ and $\Delta$  with respect to the fundamental transition, is more pronounced in the full-quantum simulation. Also, the $|a_c|^{2}$ values of the \gls{lp} and \gls{up} are beside the overall anharmonic-shift in very good agreement between the harmonic approximation and the full anharmonic treatment. Overall, we can conclude that for the coupling strengths studied, the semiclassical harmonic approximation qualitatively reproduces the main features of the vibro-polaritonic IR spectra very well, even for a very anharmonic molecule like \ce{HF}, except for the well-known general limitation of the harmonic approximation~\cite{Bloino2015-vv,Raj2022-gg}. 

Of course, the normal mode vectors $\bm{a}_i$ obtained in the harmonic approximation contain more information than only the $|a_c|^{2}$ value discussed in this section. A detailed analysis of the whole $\bm{a}_i$ vectors and a comparison withe the nuclear-photonic eigenfunctions for the case of one \ce{HF} molecule and two \ce{HF} molecules can be found in section~S3 of the supporting informations. 

\subsection{Influence of the SCF-Treatment and the Dipole Self-Energy on Vibro-Polaritonic Spectra}

\begin{figure}[htb!]
     \centering
       \includegraphics[width=0.9\textwidth]{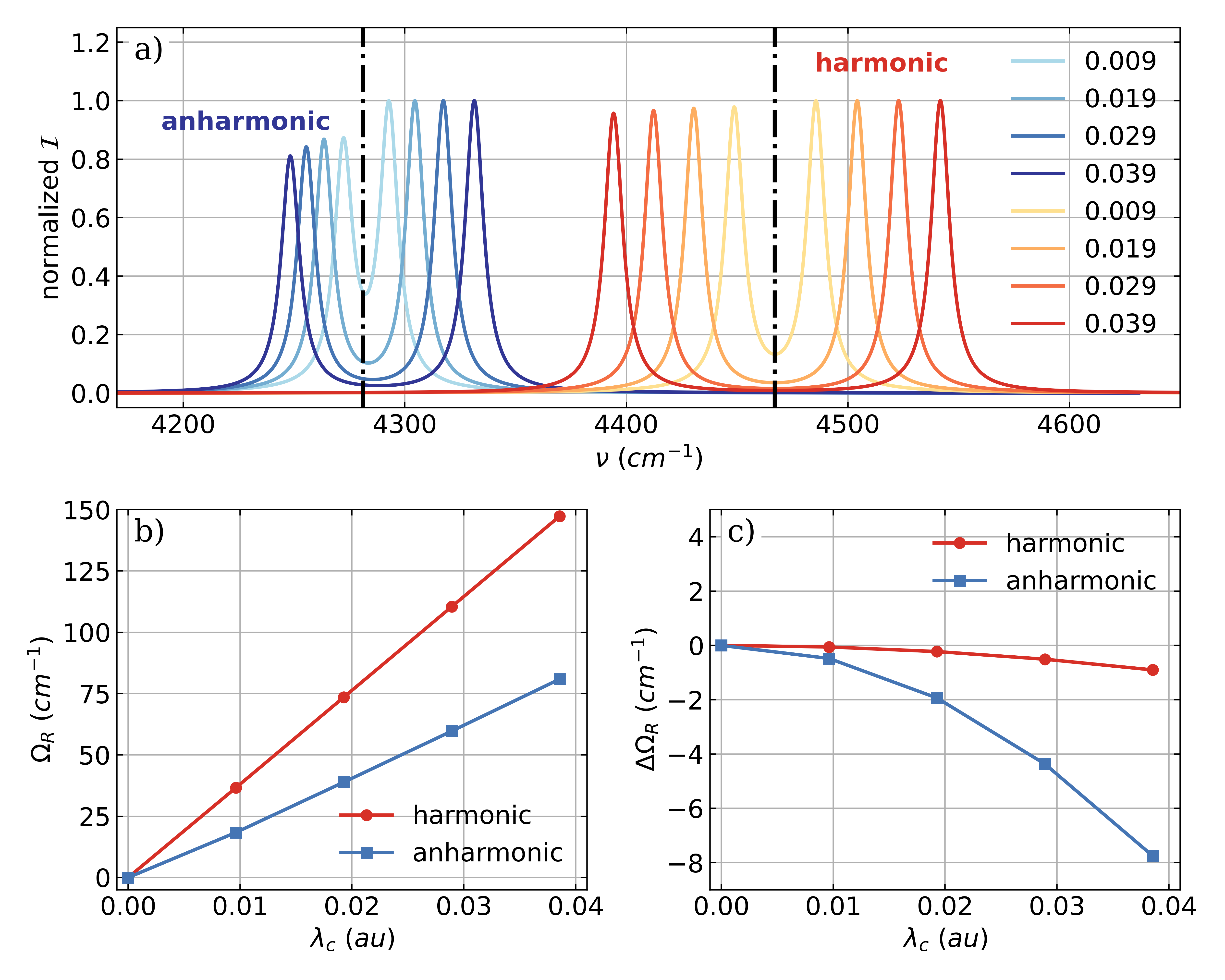}
    \caption{Vibro-polaritonic IR spectra of a single \ce{HF} molecule calculated in the harmonic approximation (redish) and in the  setup (bluish), individually normalized for each coupling strength. In both cases the full \gls{scf} treatment is neglected; for details, see text. Black dashed-dotted lines indicate the frequencies of the harmonic fundamental transition ($^{H}\nu_{1} = \SI{4467}{\per\centi\meter}$) and the anharmonic fundamental transition ($^{A}\nu_{1} = \SI{4281}{\per\centi\meter}$). The cavity frequency $\omega_c$ is resonant with the corresponding fundamental transition in both cases, and the coupling strength $\lambda_{c}$ increases from \SI{0.009}{\au} to \SI{0.039}{\au} (from light to dark color). b) Rabi splitting $\Omega_R$ as a function of $\epsilon_{c}$. c) Asymmetry $\Delta \Omega_R = \omega_c - 0.5\left( \nu^{LP} + \nu^{UP}\right)$ of the Rabi splitting.} 
\label{fig:vib_pol_spec_hf_noscf}
\end{figure}

As shown in our previous work~\cite{Sidler2023-vm,Schnappinger2023-hh}, both self-consistent treatment and the
consideration of the full \gls{dse} operator are crucial to capture relevant aspects in the description of strongly coupled molecules and molecular ensembles. Therefore, we want to determine the influence of both factors on the vibro-polaritonic IR spectra for a single \ce{HF} molecule and small ensembles with up to four \ce{HF} molecules. To analyze the influence of the \gls{scf} treatment, the vibro-polaritonic IR spectra of a single \ce{HF} molecule are calculated in the harmonic approximation and full quantum (anharmonic), neglecting the \gls{scf} of the coupled electronic photonic system. For the harmonic approximation, this is achieved by using cavity-free dipole moments, density matrix elements $D_{\alpha,\beta}$, and orbital energies $\epsilon_i$ obtained by a standard Hartree-Fock calculation to determine the \gls{cbohf} gradients and Hessians. In the full-anharmonic treatment, we use cavity-free Hartree-Fock energies and expectation values of dipole moments and \gls{dse} terms to construct the necessary potential energy surfaces in a \gls{ejcm}~\cite{Jaynes1963-os,Kowalewski2016-zo,Gudem2021-um,Couto2022-uv}. The vibro-polaritonic infrared spectra obtained for different coupling strengths calculated in the harmonic approximation and the full anharmonic as well as the trends in Rabi splitting $\Omega_R$ are shown in Fig.~\ref{fig:vib_pol_spec_hf_noscf}.

Comparing the vibro-polaritonic IR spectra without \gls{scf} (Fig.~\ref{fig:vib_pol_spec_hf_noscf}~a) with those shown in Fig.~\ref{fig:vib_pol_spec_hf}~a), both the harmonic and the full anharmonic spectra differ. Especially for the harmonic spectra, \gls{scf} has a massive influence. Without a fully converged \gls{cbohf} the Rabi splitting $\Omega_R$ is larger, see Fig.~\ref{fig:vib_pol_spec_hf_noscf}~b), and nearly symmetric with respect to the cavity frequency, see Fig.~\ref{fig:vib_pol_spec_hf_noscf}~c). Also, the intensity of the \gls{lp} transition and the \gls{up} transition are almost identical. For the full-anharmonic spectra in the \gls{ejcm} model the discrepancies seem smaller, but still clear differences can be recognized. The value of $\Omega_R$ is reproduced quite well. However, its asymmetry $\Delta \Omega_R$ is smaller and characterized by blue-shift of \gls{lp} and \gls{up} with respect to $\omega_c$. In addition, also the ratio of the intensities is inverted.  

\begin{figure}[htb!]
     \centering
         \includegraphics[width=0.8\textwidth]{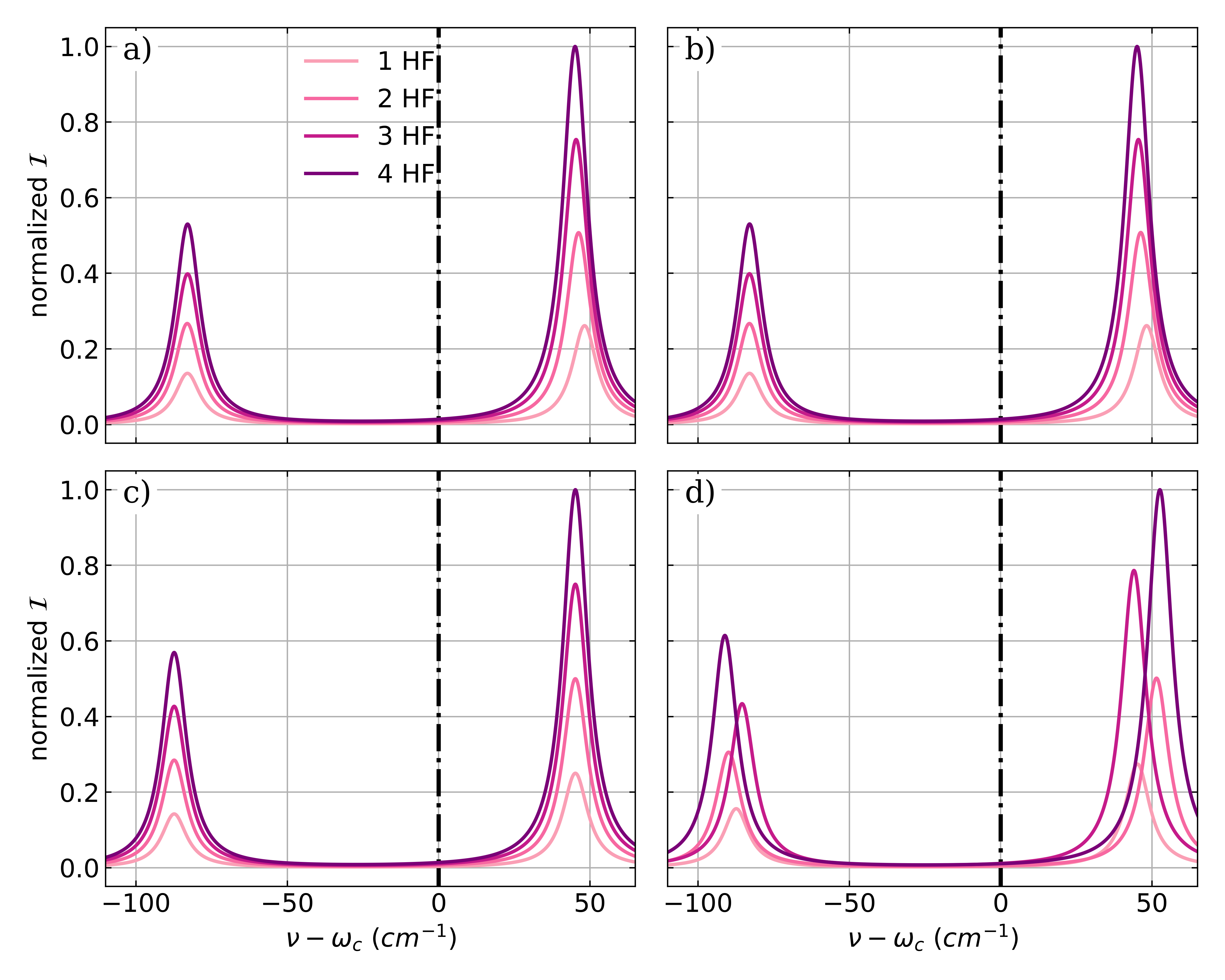}
    \caption{ Vibro-polaritonic IR spectra calculated in the harmonic approximation for different numbers of \ce{HF} molecules (color-coded) shown with respect to the cavity frequency $\omega_c$. The cavity is resonant with the harmonic fundamental transition ($^{H}\nu_{1} = \SI{4467}{\per\centi\meter}$, black dashed-dotted lines) and a rescaled coupling strength of $\lambda_{0}$ of \SI{0.057}{\au} is used (see Eq.~\eqref{eq:coupling}). a) Full \gls{cbohf} simulation in the \textit{all-parallel} configuration. b) Full \gls{cbohf} simulation in the \textit{antiparallel} configuration. c) Linear \gls{cbohf} simulation (without \gls{dse} terms) in the \textit{all-parallel} configuration. c)Linear \gls{cbohf} simulation (without \gls{dse} terms) in the \textit{antiparallel} configuration.} 
\label{fig:vib_pol_spec_hf_dse}
\end{figure}

\begin{figure}[htb!]
     \centering
         \includegraphics[width=0.8\textwidth]{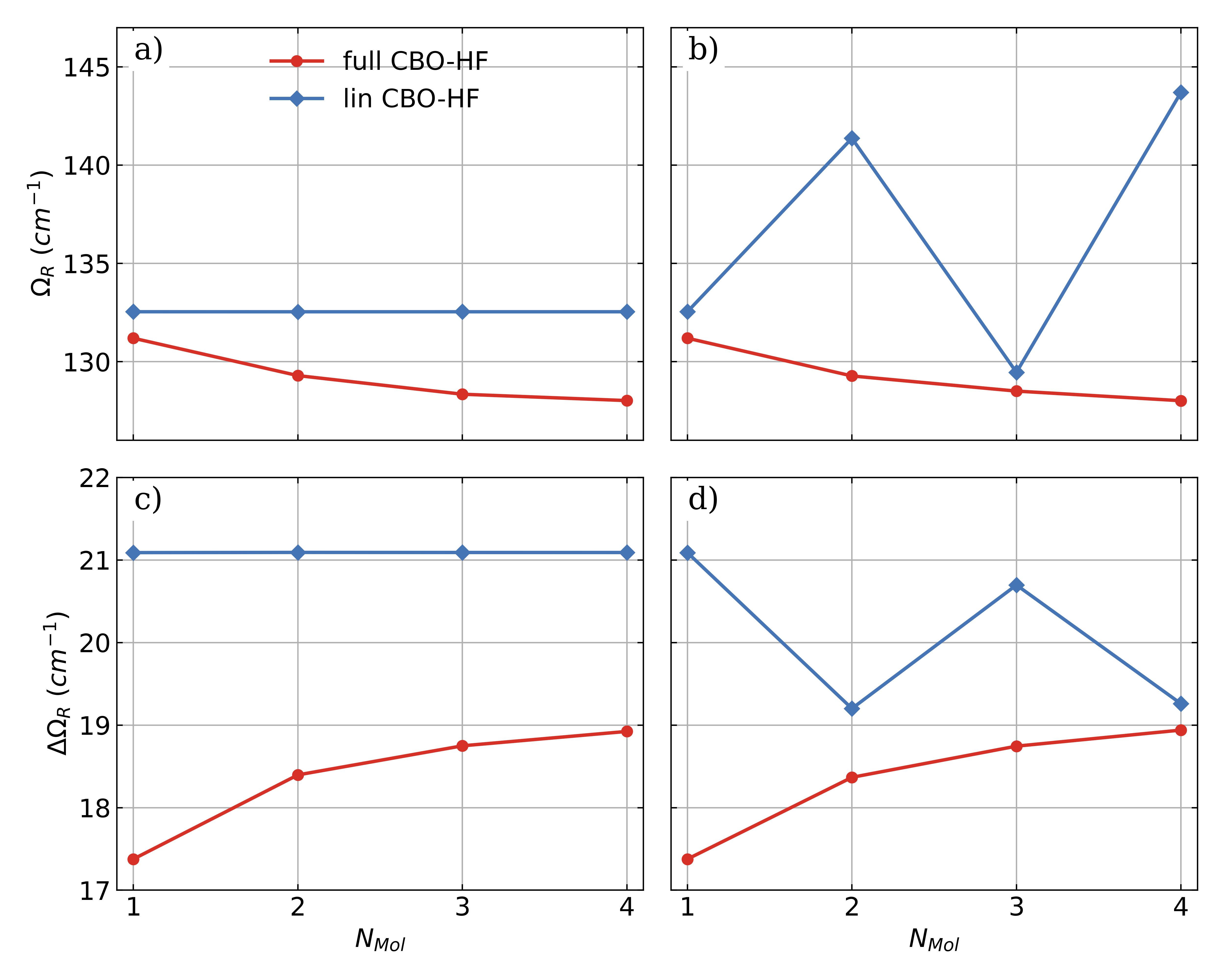}
    \caption{ Rabi splitting $\Omega_R$ and asymmetry $\Delta \Omega_R = \omega_c - 0.5\left( \nu^{LP} + \nu^{UP}\right)$ of the Rabi splitting as a function of $N_{Mol}$. Both calculated in the harmonic approximation applying full \gls{cbohf} and linear \gls{cbohf} (without \gls{dse} terms) for the \textit{all-parallel} configuration shown in a) and c) and in the \textit{antiparallel} orientation shown in b) and d). The cavity frequency $\omega_c$ is resonant with the harmonic fundamental transition (\SI{4467}{\per\centi\meter}) and a rescaled coupling strength of $\lambda_{0}$ of \SI{0.057}{\au} is used.} 
\label{fig:rabi_harm_hf_dse}
\end{figure}

Another important aspect in describing \gls{vsc} for molecules is the \gls{dse} which gives rise to a cavity-induced interaction between molecules in an ensemble and is very sensitive to the molecular orientation~\cite{Sidler2023-vm,Schnappinger2023-hh,Haugland2023-rt}. To quantify its influence on spectral features, we calculated the vibro-polaritonic IR spectra in the harmonic approximation for different numbers of \ce{HF} molecules with and without the \gls{dse} terms included in the underlying \gls{cbohf} ansatz.In the following, the case without the \gls{dse} terms is called liner \gls{cbohf}. The results for the \textit{all-parallel} and the \textit{antiparallel} configuration are shown in Fig.~\ref{fig:vib_pol_spec_hf_dse}. The observed Rabi splitting $\Omega_R$ and its asymmetry $\Delta \Omega_R$ for the \textit{all-parallel} and the \textit{antiparallel} configurations with and without the \gls{dse} term are shown as a function of the number of molecules in Fig.~\ref{fig:rabi_harm_hf_dse}

The full \gls{cbohf} spectra of the \textit{all-parallel} configuration (Fig.~\ref{fig:vib_pol_spec_hf_dse}~a)) and the \textit{antiparallel} configuration (Fig.~\ref{fig:vib_pol_spec_hf_dse}~b)) are indistinguishable. In line with our recent work~\cite{Sidler2023-vm} increasing the number of molecules ($N_{Mol}$) in the cavity while keeping the density constant (rescaling of the coupling strength, see Eq.~\eqref{eq:coupling}) leads to an increasing asymmetry $\Delta \Omega_R$ in the splitting, see Fig.~\ref{fig:rabi_harm_hf_dse}~c) and~d) red lines. At the same time, the Rabi splitting itself decreases, see Fig.~\ref{fig:rabi_harm_hf_dse}~a) and~b) red lines, although the applied rescaling should keep it constant. This may be an effect of the rather large coupling strength applied. For weaker coupling, the applied rescaling of the coupling strength leads to an almost constant value of $\Omega_R$ as the number of molecules increases, see Ref.~\citenum{Sidler2023-vm}. If the \gls{dse} contribution is neglected the two studied configurations become distinguishable, see Fig.~\ref{fig:vib_pol_spec_hf_dse}~c) and d). Comparing the \textit{all-parallel} spectra with and without \gls{dse}, the differences are rather small. The observed Rabi splitting $\Omega_R$ is slightly larger in the case without \gls{dse} and independent of the number of molecules, as shown in Fig.~\ref{fig:rabi_harm_hf_dse}~a) blue line. Furthermore, the asymmetry $\Delta \Omega_R$ is also independent of $N_{Mol}$  and constant (Fig.~\ref{fig:rabi_harm_hf_dse}~c) blue line). The observed changes are more drastic in the case of the \textit{antiparallel} configuration. In Fig.~\ref{fig:vib_pol_spec_hf_dse}~d) an oscillatory shifting of the \gls{lp} signal and \gls{up} signal is observed while changing the number of molecules. Consequently, this oscillation can be found in the Rabi splitting $\Omega_R$ and its asymmetry, see Fig.~\ref{fig:rabi_harm_hf_dse}~b)~and~d) blue line. The oscillating pattern can be attributed to the fact that odd or even
number of molecules in the \textit{antiparallel} configuration create two different situations. For even $N_{Mol}$ the whole ensemble reduces to an effective \textit{antiparallel} bimolecular case, and the situation of odd $N_{Mol}$ is equivalent to the case of a single molecule. Consistent with the literature~\cite{Rokaj2018-ww,Schafer2020-cb,Sidler2022-cg,Sidler2023-vm,Schnappinger23jctc,Schnappinger2023-hh} these results clearly show how important the \gls{dse} terms and the \gls{scf} treatment are to obtain a physically meaningful result. 

\subsection{Beyond Diatomic Molecules: Vibro-Polaritonic Spectra of \ce{NH3}}

\begin{figure}[htb!]
     \centering
    \includegraphics[width=0.9\textwidth]{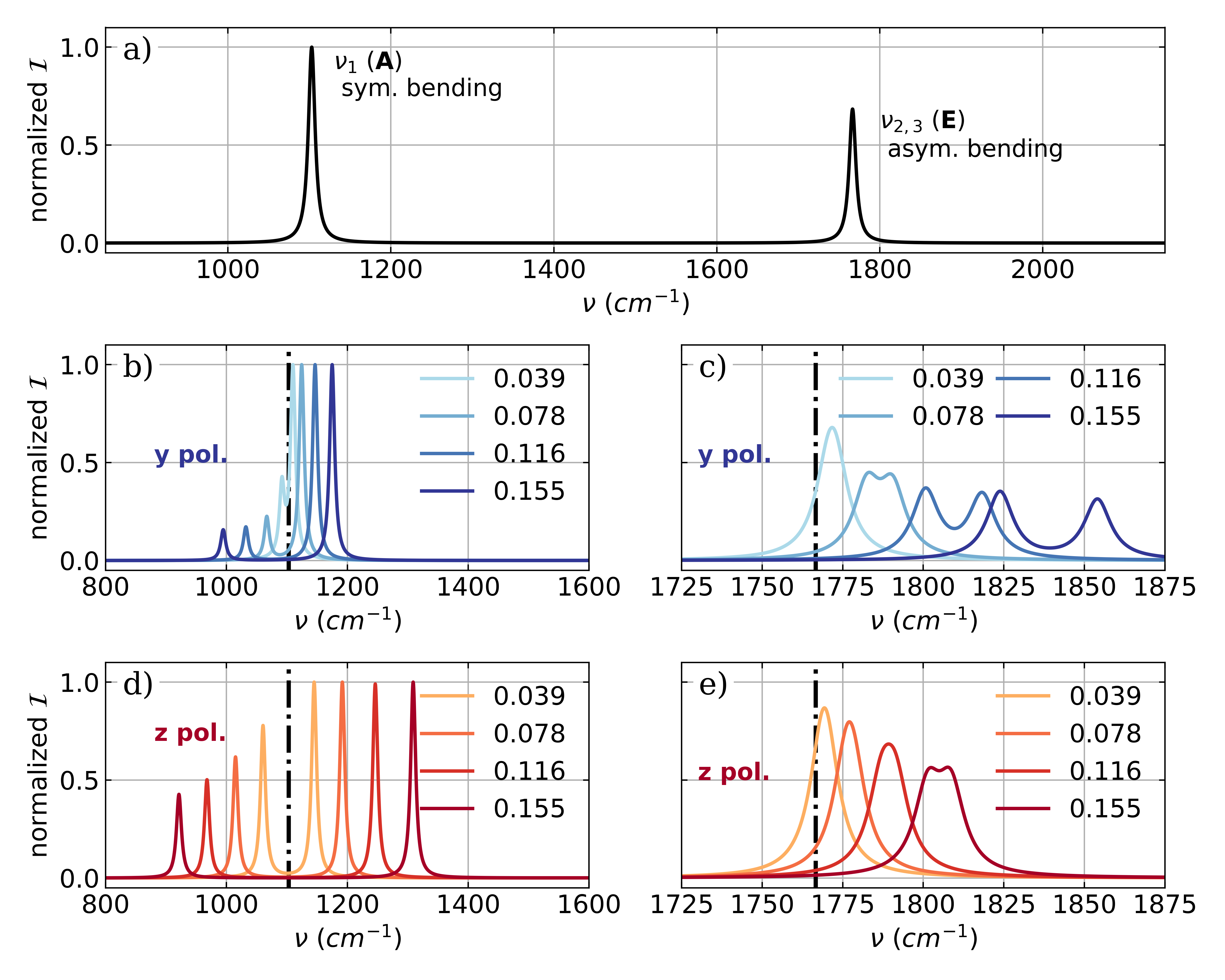}
    \caption{a) Low energy part of the vibronic IR spectra of a single \ce{NH3} molecule calculated in the harmonic approximation. The low energy part of the vibro-polaritonic IR spectra of a single \ce{NH3} molecule zoomed into the symmetric mode (b) and d)) and the two asymmetric bending modes shown in c) and e). The polarization axis of the cavity mode is the $y$ axis for b) and c) and equal to the $z$ axis for d) and e). The cavity frequency $\omega_c$ is resonant with the symmetric bending mode (\SI{1103}{\per\centi\meter}) and the cavity field strength $\lambda_{c}$ increases from \SI{0.039}{\au} to \SI{0.155}{\au}.} 
\label{fig:spectrum_nh3_v1_bend}
\end{figure}

So far we have discussed only diatomic molecules, which have only one vibrational degree of freedom, or, in the case of the ensembles, linear combinations of the same vibrational degree. However, since we have validated the harmonic approximation and the associated normal mode analysis, its semi-classical nature also allows us to treat larger molecular systems. Therefore, in the last part of this work we will study the vibro-polaritonic IR spectra of a single \ce{NH3} molecule. The bare \ce{NH3} molecule has six vibrational degrees of freedom: One symmetric bending mode ($\nu_1$), two degenerate asymmetric bending modes ($\nu_{2,3}$), one symmetric stretching mode ($\nu_4$) and two degenerate asymmetric stretching modes ($\nu_{5,6}$). Since \ce{NH3} is not a linear molecule and the vibrational modes have different spatial orientation (symmetry), the polarization axis of the single cavity mode plays a significant role. In our simulations, the single \ce{NH3} molecule oriented with respect to the center of nuclear charges and the lone pair orbital of nitrogen is pointing along the $z$ axis of the laboratory frame. Since for this chosen molecular orientation the $x$ axis and the $y$ axis are equivalent, only the $y$ polarization and the $z$ polarization of the single cavity mode are discussed in the following. The cavity frequency $\omega_c$ is set to be resonant with the symmetric bending mode $\nu_1$, which has the lowest energy of \SI{1103}{\per\centi\meter} and is the strongest transition in the cavity-free case. The bare molecular vibronic IR spectrum and the vibro-polaritonic IR spectra of a single \ce{NH3} molecule are shown in Fig.~\ref{fig:spectrum_nh3_v1_bend} for the bending modes and in Fig.~S14 of the supporting information for the stretching modes.

As expected, the splitting into a \gls{lp} and a \gls{up} transition is observed for the symmetric bending mode $\nu_{1}$ when coupled to a resonant optical cavity, see Fig.~\ref{fig:spectrum_nh3_v1_bend}~b) and~d). The Rabi splitting $\Omega_R$ increases with increasing coupling strength and is larger for $z$ polarization than for $x/y$ polarization due to the chosen molecular orientation. Interestingly, the signal of the asymmetric bending modes ($\nu_{2}$ and $\nu_{3}$) is also affected by the coupling to the cavity mode, as shown in Fig.~\ref{fig:spectrum_nh3_v1_bend}~c) and~e). For the $x/y$ polarization the signal is blue-shifted and for a coupling  strength $\lambda_{c}$ larger than \SI{0.078}{\au} the degeneracy of the two transitions is lifted. This shift and the splitting into two separate signals is also observed for the $z$ polarization. However, it is smaller for this polarization and occurs only at higher coupling strengths.

\begin{figure}[htb!]
     \centering
         \includegraphics[width=0.9\textwidth]{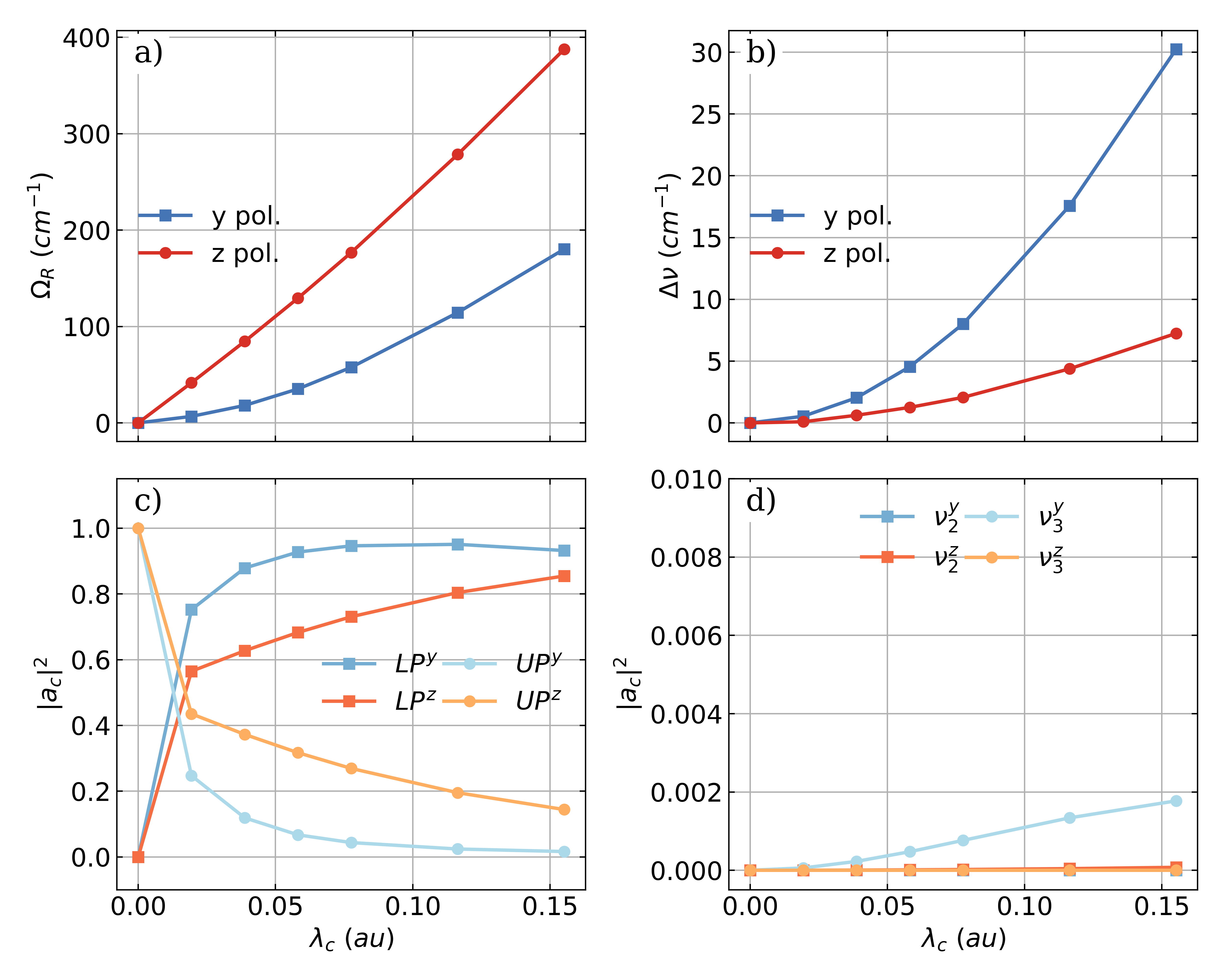}
    \caption{a) Rabi splitting $\Omega_R$ for the symmetric bending mode of a single \ce{NH3} molecule as a function of $\lambda_{c}$. b) Splitting $\Delta \nu$ of the asymmetric bending modes due to cavity interaction. 
    The normal mode value $|a_c|^{2}$ describing the change in $q_c$ for the \gls{up} and \gls{lp} modes c) and for the two asymmetric bending modes d). The cavity frequency $\omega_c$ is resonant with the symmetric bending mode (\SI{1103}{\per\centi\meter}) and both $y/x$ (bluish) and $z$ (reddish) polarization of the cavity mode are shown.} 
\label{fig:splitting_nh3_v1_bending}
\end{figure}

To better understand and characterize the two observed splittings in the vibro-polaritonic IR spectra of \ce{NH3}, their magnitude is plotted for both polarization directions as a function of the coupling strength $\lambda_{c}$ in Fig.~\ref{fig:splitting_nh3_v1_bending}~a) and~b). Furthermore, the corresponding values $|a_c|^{2}$ for the associated normal modes for both splittings are plotted in Fig.~\ref{fig:splitting_nh3_v1_bending}~c) and d) as a function of $\lambda_{c}$. These $|a_c|^{2}$ values describe the change in the classical photon displacement field $q_c$ and probes how photonic the corresponding vibrational transition is.

Clearly, the formation of a \gls{lp} and a \gls{up} transition for the symmetric bending mode is more efficient for the $z$ polarization since the observed Rabi splitting $\Omega_R$ is larger and scales almost linear with $\epsilon_{c}$, see  Fig.~\ref{fig:splitting_nh3_v1_bending}~a). For the $y/x$ polarization the value is smaller but the scaling is close to quadratic. The stronger effect observed for $z$ polarization is understandable, since the symmetric bending mode mainly involves motion along the $z$ axis in the laboratory frame and only very small changes along the $y/x$ direction. A closer look at the $|a_c|^{2}$ values for the \gls{lp} and \gls{up} normal modes (Fig.~\ref{fig:splitting_nh3_v1_bending}~c) shows a similar behavior to the case of the \ce{HF} molecule. Once the cavity coupling is present, the $|a_c|^{2}$ values are clearly non-zero for both \gls{lp} and \gls{up}. Since for the $z$ polarization the coupling between the cavity mode and the symmetric bending mode is more efficient, the $|a_c|^{2}$ value of the \gls{up} mode is closer to that of the \gls{lp} mode and clearly larger compared to the $y/x$ case. In other words, the light and the molecule are less hybridized for the $y/x$ polarization. The splitting and the $|a_c|^{2}$ values for the two asymmetric bending modes shown in Fig.~\ref{fig:splitting_nh3_v1_bending}~b) and d) behave differently. In the $z$ polarization, where a smaller splitting is observed, both $|a_c|^{2}$ values are lower than $0.001$. Even for the $y/x$ case, they do not become larger than $0.002$, which is still orders of magnitude smaller than for the \gls{lp} and the \gls{up} transition. This is clear evidence that both asymmetric bending modes mostly are matter transitions and no light-matter hybrid states are formed. The observed splitting of $\nu_2$ and $\nu_3$ when coupled to a cavity can be explained by a symmetry breaking by the confined photon mode. Its polarization axis in $x$ or $y$ makes these two directions, which are identical in the case without cavity, distinguishable. Note that in our current simulations we only consider a single photon mode, and in a more general setup, there are many modes with usually more than one polarization direction. Therefore, we assume that the observed splitting due to symmetry breaking is overestimated in our calculations. However, the fact that the interaction with an optical cavity not only affects near-resonance transitions by forming \gls{lp} and \gls{up} states, but also modifies all signals by slightly changing the energetics and breaking molecular symmetries is probably still important in more general cases and already observed earlier in literature~\cite{Bonini2022-ee}.

\section{Summary and Conclusion}

Based on the recently formulated cavity Born-Oppenheimer Hartree-Fock ansatz~\cite{Schnappinger2023-hh} we have introduced a wave function-based methodology to calculate the vibro-polaritonic IR spectra in a \textit{ab-initio} manner. By applying the well-known harmonic approximation, we have access to the vibrational frequencies and normal modes of systems when light and matter are strongly coupled. Using the cavity Born-Oppenheimer approximation, the obtained normal modes combine both nuclear and photonic degrees of freedom, allowing a detailed analysis of the vibro-polariton states. The necessary second derivatives of the \gls{cbohf} energy are calculated via finite differences of the analytic first derivatives (gradients) introduced in this work. 

We demonstrate the capability of our framework by carefully comparing the vibro-polaritonic IR spectra obtained for small ensembles of \ce{HF} molecules with full quantum-mechanical calculations. Overall, the semi-classical harmonic approximation qualitatively reproduces the main features of the vibro-polaritonic IR spectra very well. It even captures to some extent polarization effects such as cavity detuning previously observed~\cite{Sidler2023-vm} when the full \gls{dse} contribution is included and the coupled electronic-photonic problem is solved self-consistently. The semi-classical nature of the harmonic approximation allows efficient description of vibro-polaritonic IR spectra of large molecular systems or small ensembles. As a first test case, we simulated the vibro-polaritonic spectra of a single \ce{NH3} molecule. In addition to the expected formation of vibro-polaritonic states, the coupling to an optical cavity changes the whole vibrational spectrum. The presence of the confined light mode leads to an energy shift of most signals and, depending on the cavity polarization and the molecular orientation, the molecular symmetry is reduced.

The new analytical gradients used here to obtain vibro-polaritonic IR spectra open many avenues for future exploration. The gradients and the numerical Hessian can be used to optimize the molecular system coupled to an optical cavity. In contrast to the cavity/field-free case, rotation with respect to the laboratory frame is no longer trivial for optimizations in the presence of a cavity mode with a defined polarization axis~\cite{Schnappinger2023-hh}. Beside the possibility to optimize molecules under \gls{vsc}, the analytical gradients can be used to perform \textit{ab-initio} semi-classical dynamics simulations for single molecules or small ensembles.

\begin{acknowledgement}
We thank Dominik Sidler and Michael Ruggenthaler for inspiring discussions and helpful comments. M.K. acknowledges funding from the European Union’s Horizon 2020 research and innovation program under the Marie Skłodowska-Curie grant agreement No. 860553. This project has received funding from the European Research Council (ERC) under the European Union’s Horizon 2020 research and innovation program (grant agreement no. 852286).
\end{acknowledgement}

\begin{suppinfo}
See the supplementary material for the details of the derivation of the \gls{cbohf} energy gradients,  additional figures for the discussion of \ce{HF} and \ce{NH3} coupled to an optical cavity and a detailed comparison of nuclear-photonic eigenfunctions and vibro-polaritonic normal modes. All data underlying this study are available from the corresponding author upon reasonable request.
\end{suppinfo}

\bibliography{lit.bib}
\providecommand{\latin}[1]{#1}
\makeatletter
\providecommand{\doi}
  {\begingroup\let\do\@makeother\dospecials
  \catcode`\{=1 \catcode`\}=2 \doi@aux}
\providecommand{\doi@aux}[1]{\endgroup\texttt{#1}}
\makeatother
\providecommand*\mcitethebibliography{\thebibliography}
\csname @ifundefined\endcsname{endmcitethebibliography}
  {\let\endmcitethebibliography\endthebibliography}{}

\end{document}


\maketitle

\tableofcontents
\clearpage
\section{Derivation of the CBO-Hartree-Fock gradients}

The energy expectation value $\bigl\langle  E_{CBO} \bigr\rangle$ in the basis of the atomic orbitals has the following form:
\begin{equation}
\label{eq:Ecbo}
\begin{split}
\bigl\langle  E_{CBO} \bigr\rangle &=  \sum_{\alpha,\beta}^{M}  D_{\alpha,\beta}  \Biggl( \bigl\langle  \alpha  \big| \hat{h} \big| \beta \bigr\rangle   + \omega_c q_{c}
\bigl\langle \alpha \big| \hat{x} \big| \beta \bigr\rangle +  
X  \bigl\langle \alpha \big| \hat{x} \big| \beta \bigr\rangle
+ \frac{1}{2} \bigl\langle \alpha \big| \hat{x}^2 \big| \beta \bigr\rangle \Biggr) \\
&+  \frac{1}{2} \sum_{\alpha,\beta,\gamma,\delta}^{M} D_{\alpha,\beta} D_{\gamma,\delta} \Biggl( \bigl\langle \alpha  \gamma \big| \hat{g} \big| \beta \delta \bigr\rangle 
 + \bigl\langle \alpha  \big| \hat{x} \big| \beta \bigr\rangle  \bigl\langle \gamma \big| \hat{x} \big| \delta \bigr\rangle 
 - \bigl\langle \alpha  \big| \hat{x} \big| \delta \bigr\rangle  \bigl\langle \gamma \big| \hat{x} \big| \beta \bigr\rangle  \Biggr) \\
 & + \tilde{E}_{n,c} \quad  \text{ with} \quad \tilde{E}_{n,c} =  V_{nn} + \frac{1}{2} \omega_c^2 q_{c}^2 +  E^{(nuc)}_{lin} + E^{(nuc)}_{dse} .  
  \end{split}
\end{equation}
Where $\hat{x}$ and $X$  are the electronic and nuclear part of the projected molecular dipole moment:
\begin{equation}
\hat{x}=  -\bm{\lambda}_{c} \cdot \bm{\hat{r}} \  \text{ and} \ X = \bm{\lambda}_{c} \cdot \left(  \sum_{A=1}^{N_{Nuc}} Z_{A} \bm{R}_A \right)
\end{equation}
The first derivative of the energy $\bigl\langle  E_{CBO} \bigr\rangle$ with respect to a nuclear or photon displacement coordinate $\zeta_i$ is
\begin{equation}
\label{eq:def_grad}
   \frac{\partial}{\partial \zeta_i} \bigl\langle  E_{CBO} \bigr\rangle  =  \sum_{\alpha,\beta}^{M}  D_{\alpha,\beta}  \frac{\partial}{\partial \zeta_i} F_{\alpha,\beta} + \sum_{\alpha,\beta}^{M} F_{\alpha,\beta}  \frac{\partial}{\partial \zeta_i} D_{\alpha,\beta}   + \frac{\partial}{\partial \zeta_i} \tilde{E}_{n,c}.
\end{equation}
The first term in Eq.~\eqref{eq:def_grad} is the Hellmann-Feynman term, the second term is the wavefunction derivative, or Pulay term, and the last part is the derivative of all scalar energy contributions. The full Hellmann-Feynman term has the following structure:
\begin{equation}
\label{eq:dev_HF}
\begin{split}
\sum_{\alpha,\beta}^{M}  D_{\alpha,\beta}  \frac{\partial}{\partial \zeta_i} F_{\alpha,\beta}  = & \sum_{\alpha,\beta}^{M}  D_{\alpha,\beta}  \frac{\partial}{\partial \zeta_i} \Biggl( \bigl\langle  \alpha  \big| \hat{h} \big| \beta \bigr\rangle   + \omega_c q_{c}
\bigl\langle \alpha \big| \hat{x} \big| \beta \bigr\rangle +  
X  \bigl\langle \alpha \big| \hat{x} \big| \beta \bigr\rangle
+ \frac{1}{2} \bigl\langle \alpha \big| \hat{x}^2 \big| \beta \bigr\rangle \Biggr) \\
&+  \frac{1}{2} \sum_{\alpha,\beta,\gamma,\delta}^{M} D_{\alpha,\beta} D_{\gamma,\delta} \frac{\partial}{\partial \zeta_i} \Biggl(\bigl\langle \alpha  \gamma \big| \hat{g} \big| \beta \delta \bigr\rangle 
 + \bigl\langle \alpha  \big| \hat{x} \big| \beta \bigr\rangle  \bigl\langle \gamma \big| \hat{x} \big| \delta \bigr\rangle 
 - \bigl\langle \alpha  \big| \hat{x} \big| \delta \bigr\rangle  \bigl\langle \gamma \big| \hat{x} \big| \beta \bigr\rangle  \Biggr)
  \end{split}
\end{equation}
Since the \gls{cbohf} wave function is variational optimized, the explicit calculation of the density derivatives can be avoided~\cite{Pulay1969-hg} and the Pulay term can be written in terms of overlap integral derivatives:
\begin{equation}
\label{eq:dev_density}
\begin{split}
        \sum_{\alpha,\beta}^{M} F_{\alpha,\beta} \frac{\partial D_{\alpha,\beta} }{\partial \zeta_i}  &= - \sum_{\alpha,\beta}^{M} W_{\alpha,\beta}  \frac{\partial S_{\alpha,\beta} }{\partial \zeta_i}\\
        &= \sum_{\alpha,\beta}^{M}  \sum_{i=1}^{N_{oc}} \epsilon_i c_{i,\alpha}^{*}c_{i,\beta} \left( 
\bigl\langle \frac{\partial \alpha}{\zeta_i}  \big| \beta \bigr\rangle +\bigl\langle \alpha \big| \frac{\partial \beta}{\zeta_i} \bigr\rangle  \right)
\end{split}
\end{equation}
The remaining derivative is simpler since it dose not involve electron coordinates:
\begin{equation}
\label{eq:dev_nuc}
    \frac{\partial}{\partial \zeta_i} \tilde{E}_{n,c} = \frac{\partial}{\partial \zeta_i} \Biggl( V_{nn} + \frac{1}{2} \omega_c^2 q_{c}^2 +  E^{(nuc)}_{lin} + E^{(nuc)}_{dse} \Biggr)
\end{equation}

In the following two cases will be discussed, the first where $\zeta_i$ is a nuclear coordinate $R_{i}$ and the second where $\zeta_i$ is a photon displacement coordinate $q_c$. For the first case, the nuclear derivatives of $\bigl\langle\hat{h}\bigl\rangle$, $\bigl\langle\hat{g}\bigl\rangle$ (in Eq.~\eqref{eq:dev_HF}), $V_{nn}$ (in Eq.~\eqref{eq:dev_nuc}) and the overlap integral derivatives in (in Eq.~\eqref{eq:dev_density}) are identical to the standard Hartree-Fock gradient terms and can be found in the literature~\cite{Pulay1969-hg,Pople2009-ho,Jensen2017-fq,aszabo82-qc}. Therefore, we focus on the "new" terms introduced by the \gls{cbohf} ansatz in Eq.~\eqref{eq:dev_HF} and Eq.~\eqref{eq:dev_nuc}.
The nuclear derivative of the linear cavity-electron interaction term  takes the following form:
\begin{equation}
\begin{split}
\frac{\partial}{\partial R_{i}} \omega_c q_{c}
\bigl\langle \alpha \big| \hat{x} \big| \beta \bigr\rangle & = \omega_c q_{c} \left( 
\bigl\langle \frac{\partial \alpha}{\partial R_{i}}  \big| \hat{x}\big| \beta \bigr\rangle + \bigl\langle \alpha \big| \frac{\partial \hat{x}}{\partial R_{i}} \big| \beta \bigr\rangle  + \bigl\langle \alpha \big| \hat{x} \big| \frac{\partial \beta}{\partial R_{i}} \bigr\rangle  \right) \\
& = \omega_c q_{c} \left( 
\bigl\langle \frac{\partial \alpha}{\partial R_{i}}  \big| \hat{x}\big| \beta \bigr\rangle +\bigl\langle \alpha \big| \hat{x} \big| \frac{\partial \beta}{\partial R_{i}} \bigr\rangle  \right)
\end{split}
\end{equation}
Here the central or Hellmann-Feynman term becomes equal zero since $\hat{x}$ is independent of $R_{i}$. The nuclear derivatives of the pure electronic contributions to the \gls{dse} follow a similar scheme. Whereby the one-electron contributions (Eq.~\eqref{eq:dev_dse1}) are simpler than the two electron contributions (Eq.~\eqref{eq:dev_dse2j}).
\begin{equation}
\label{eq:dev_dse1}
\frac{1}{2} \frac{\partial}{\partial R_{i}}   \bigl\langle \alpha \big| \hat{x}^2 \big| \beta \bigr\rangle = \frac{1}{2} \left( 
\bigl\langle \frac{\partial \alpha}{\partial R_{i}}  \big| \hat{x}^2 \big| \beta \bigr\rangle +\bigl\langle \alpha \big| \hat{x}^2 \big| \frac{\partial \beta}{\partial R_{i}} \bigr\rangle  \right)
\end{equation}
\begin{equation}
\label{eq:dev_dse2j}
\begin{split}
\frac{1}{2} \frac{\partial}{\partial R_{i}} \bigl\langle \alpha  \big| \hat{x} \big| \beta \bigr\rangle  \bigl\langle \gamma \big| \hat{x} \big| \delta \bigr\rangle &= \frac{1}{2} \left( \frac{\partial}{\partial R_{i}} \bigl\langle \alpha  \big| \hat{x} \big| \beta \bigr\rangle \right) \bigl\langle \gamma \big| \hat{x} \big| \delta \bigr\rangle + \frac{1}{2} \bigl\langle \alpha  \big| \hat{x} \big| \beta \bigr\rangle \left(\frac{\partial}{\partial R_{i}}  \bigl\langle \gamma \big| \hat{x} \big| \delta \bigr\rangle \right)\\
&= \frac{1}{2} \left( \bigl\langle \frac{\partial \alpha}{\partial R_{i}}  \big| \hat{x}\big| \beta \bigr\rangle +\bigl\langle \alpha \big| \hat{x} \big| \frac{\partial \beta}{\partial R_{i}} \bigr\rangle \right)\bigl\langle \gamma \big| \hat{x} \big| \delta \bigr\rangle + \frac{1}{2} \bigl\langle \alpha  \big| \hat{x} \big| \beta \bigr\rangle  \left( 
\bigl\langle \frac{\partial \gamma}{\partial R_{i}}  \big| \hat{x}\big| \delta  \bigr\rangle +\bigl\langle\gamma \big| \hat{x} \big| \frac{\partial \delta }{\partial R_{i}} \bigr\rangle \right)
\end{split}
\end{equation}
The nuclear derivative for the mixed electron-nuclear \gls{dse} is 
\begin{equation}
\begin{split}
\frac{\partial}{\partial R_{i}}  X  \bigl\langle \alpha \big| \hat{x} \big| \beta \bigr\rangle &= \left( \frac{\partial}{\partial R_{i}}  X \right) \bigl\langle \alpha \big| \hat{x} \big| \beta \bigr\rangle + X \left( \frac{\partial}{\partial R_{i}} \bigl\langle \alpha \big| \hat{x} \big| \beta \bigr\rangle \right) \\
&= \left( \frac{\partial}{\partial R_{i}}  \bm{\lambda}_{c} \cdot \left(  \sum_{A=1}^{N_{Nuc}} Z_{A} \bm{R}_A \right) \right) \bigl\langle \alpha \big| \hat{x} \big| \beta \bigr\rangle + X \left( 
\bigl\langle \frac{\partial \alpha}{\partial R_{i}}  \big| \hat{x}\big| \beta \bigr\rangle +\bigl\langle \alpha \big| \hat{x} \big| \frac{\partial \beta}{\partial R_{i}} \bigr\rangle  \right) \\
&= \lambda_{i} Z_{i} \bigl\langle \alpha \big| \hat{x} \big| \beta \bigr\rangle + X \left( 
\bigl\langle \frac{\partial \alpha}{\partial R_{i}}  \big| \hat{x}\big| \beta \bigr\rangle +\bigl\langle \alpha \big| \hat{x} \big| \frac{\partial \beta}{\partial R_{i}} \bigr\rangle  \right)
\end{split}
\end{equation}
Regarding Eq.~\eqref{eq:dev_nuc} there are only two "new" nuclear derivatives:
\begin{equation}
\frac{\partial}{\partial R_{i}} E^{(nuc)}_{lin} = \frac{\partial}{\partial R_{i}} \omega_c  q_c  X = \frac{\partial}{\partial R_{i}} \omega_c  q_c  \bm{\lambda}_{c} \cdot \left(  \sum_{A=1}^{N_{Nuc}} Z_{A} \bm{R}_A \right) =   \omega_c  q_c  \lambda_{i} Z_{i} 
\end{equation}
\begin{equation}
\begin{split}
\frac{\partial}{\partial R_{i}} E^{(nuc)}_{dse}  = \frac{1}{2} \frac{\partial}{\partial R_{i}}  \left( \bm{\lambda}_{c} \cdot \bm{\mu}_{Nuc} \right)^2 = \frac{1}{2} \frac{\partial}{\partial R_{i}}  \left(  \bm{\lambda}_{c} \cdot \left(  \sum_{A=1}^{N_{Nuc}} Z_{A} \bm{R}_A \right)  \right)^2 
\end{split}
\end{equation}

For the second case $\zeta_i$ being a photon displacement coordinate $q_c$ the Pulay term in  Eq.~\eqref{eq:dev_density} is zero, since the atomic orbitals used are independent of $q_c$. This is also the reason why for the Hellmann-Feynman part of the derivative (Eq.~\eqref{eq:dev_HF}) only terms that explicitly depend on $q_c$ are of relevance. In combination with relevant parts of Eq.~\eqref{eq:dev_nuc} the following expressions for the derivative with respect to $q_c$ is found:
\begin{equation}
\begin{split}
 \frac{\partial}{\partial q_c} E_{CBO} =& \omega_c^2 q_c  + \omega_c \sum_{\alpha,\beta}^{M} D_{\alpha,\beta}  \bigl\langle \alpha \big| \hat{x} \big| \beta \bigr\rangle  + \omega_c  \bm{\lambda}_{c} \cdot  \bm{\mu}_{Nuc}  \\
 =&  \omega_c^2 q_c  - \omega_c \left( \bm{\lambda}_{c} \cdot  \bigl\langle \bm{\hat{\mu}} \bigr\rangle \right)
 \end{split}
 \end{equation}
This equivalent to the result obtained using the Hellmann–Feynman theorem~\cite{Bonini2022-ee,Schnappinger2023-hh}.
\clearpage

\section{Validation of the Harmonic Approximation}

\begin{figure}[htb!]
     \centering
         \includegraphics[width=0.8\textwidth]{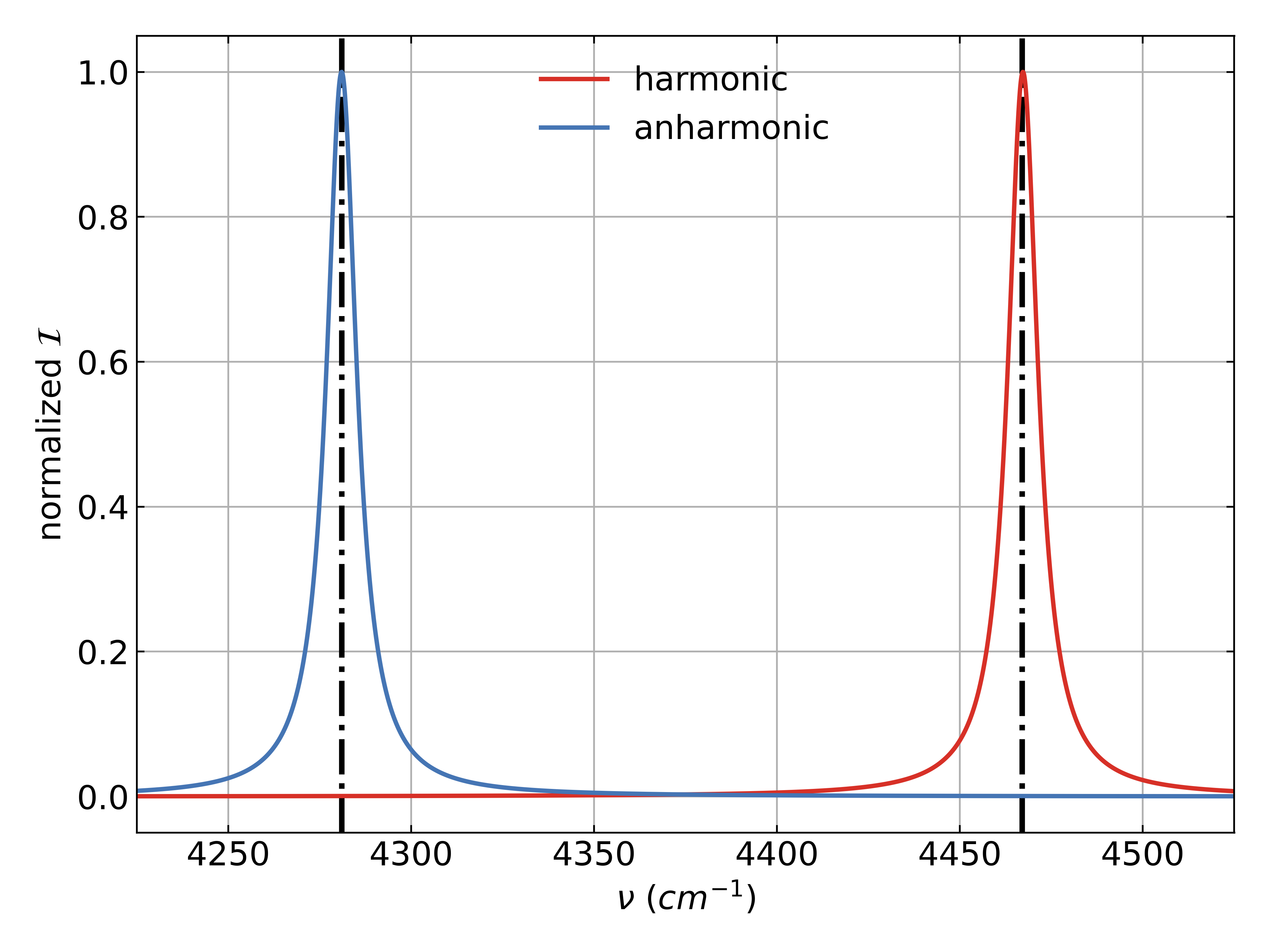}
    \caption{Vibronic IR spectra of a single \ce{HF} molecule calculated in the harmonic approximation (red) and the full anharmonic simulation (blue). Black dashed-dotted lines indicate the frequencies of the harmonic (\SI{4467}{\per\centi\meter}) and anharmonic (\SI{4281}{\per\centi\meter}) fundamental transitions.} 
\label{fig:vib_spec_hf}
\end{figure}

The spectra of an individual \ce{HF} molecule without coupling to an optical cavity calculated in the harmonic approximation and full quantum mechanical (anharmonic) are shown in Fig.~\ref{fig:vib_spec_hf}. The fundamental vibrational transition in the harmonic approximation has a frequency of $^{H}\nu_{1} = \SI{4467}{\per\centi\meter}$. The fully quantum mechanical treatment leads to a strong red-shifted of \SI{186}{\per\centi\meter} for the fundamental vibrational transition ($^{A}\nu_{1} = \SI{4281}{\per\centi\meter}$).

\begin{figure}[htb!]
     \centering
         \includegraphics[width=0.9\textwidth]{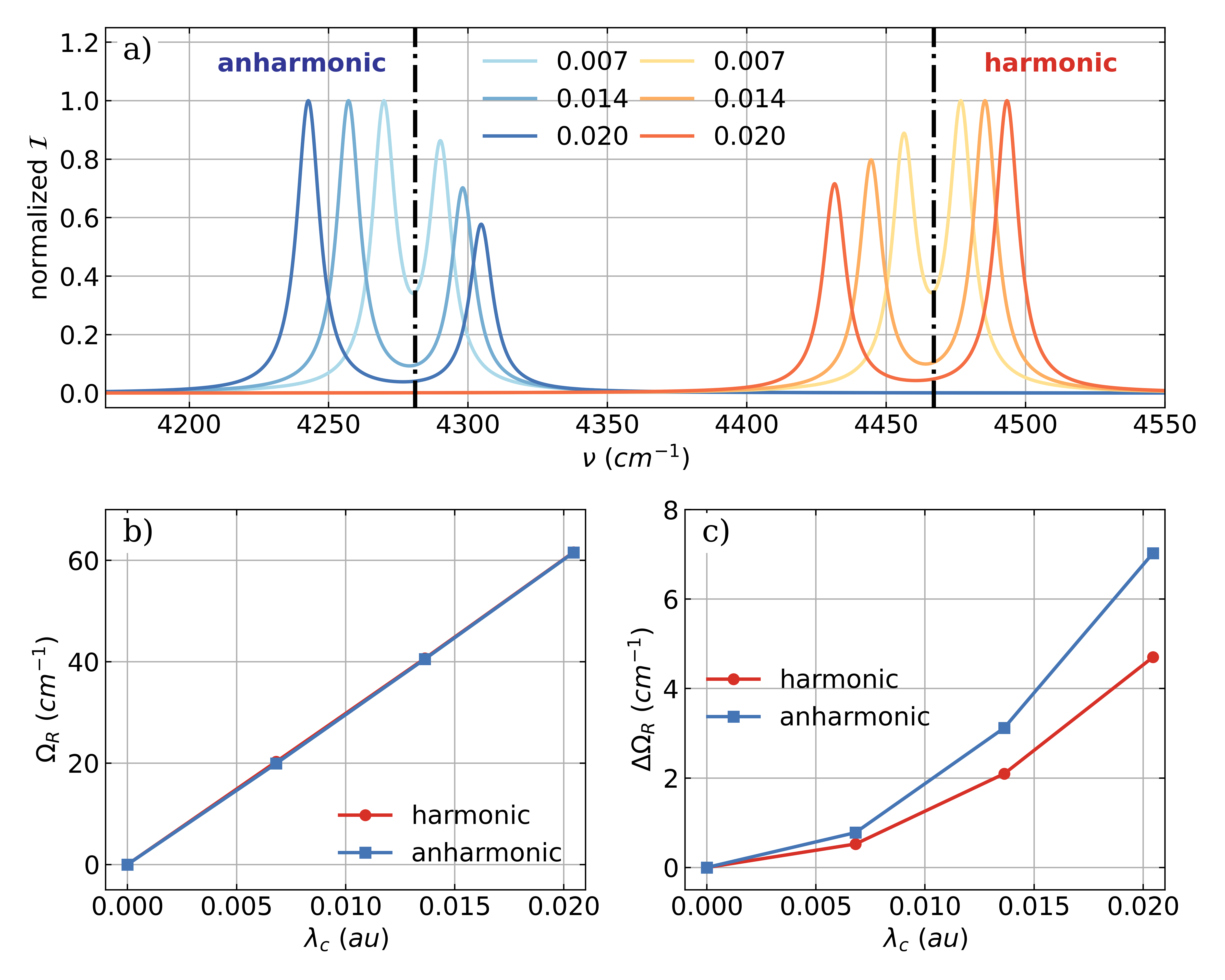}
    \caption{a) Vibro-polaritonic IR spectra of a two parallel \ce{HF} molecule calculated in the harmonic approximation (redish) and in the full anharmonic setup (bluish). Black dashed-dotted lines indicate the frequencies of the harmonic (\SI{4467}{\per\centi\meter}) and anharmonic (\SI{4281}{\per\centi\meter}) fundamental transitions. The cavity frequency $\omega_c$ is resonant with the corresponding fundamental transition in both cases, and the coupling strength $\lambda_{c}$ is increased from \SI{0.009}{\au} to \SI{0.039}{\au} (from lightest to darkest color). b) Rabi splitting $\Omega_R$ as a function of $\epsilon_{c}$. c) Asymmetry $\Delta \Omega_R = \omega_c - 0.5\left( \nu^{LP} + \nu^{UP}\right)$ of the Rabi splitting.} 
\label{fig:spectrum_2p_hf_resonant}
\end{figure}

\begin{figure}[htb!]
     \centering
         \includegraphics[width=0.9\textwidth]{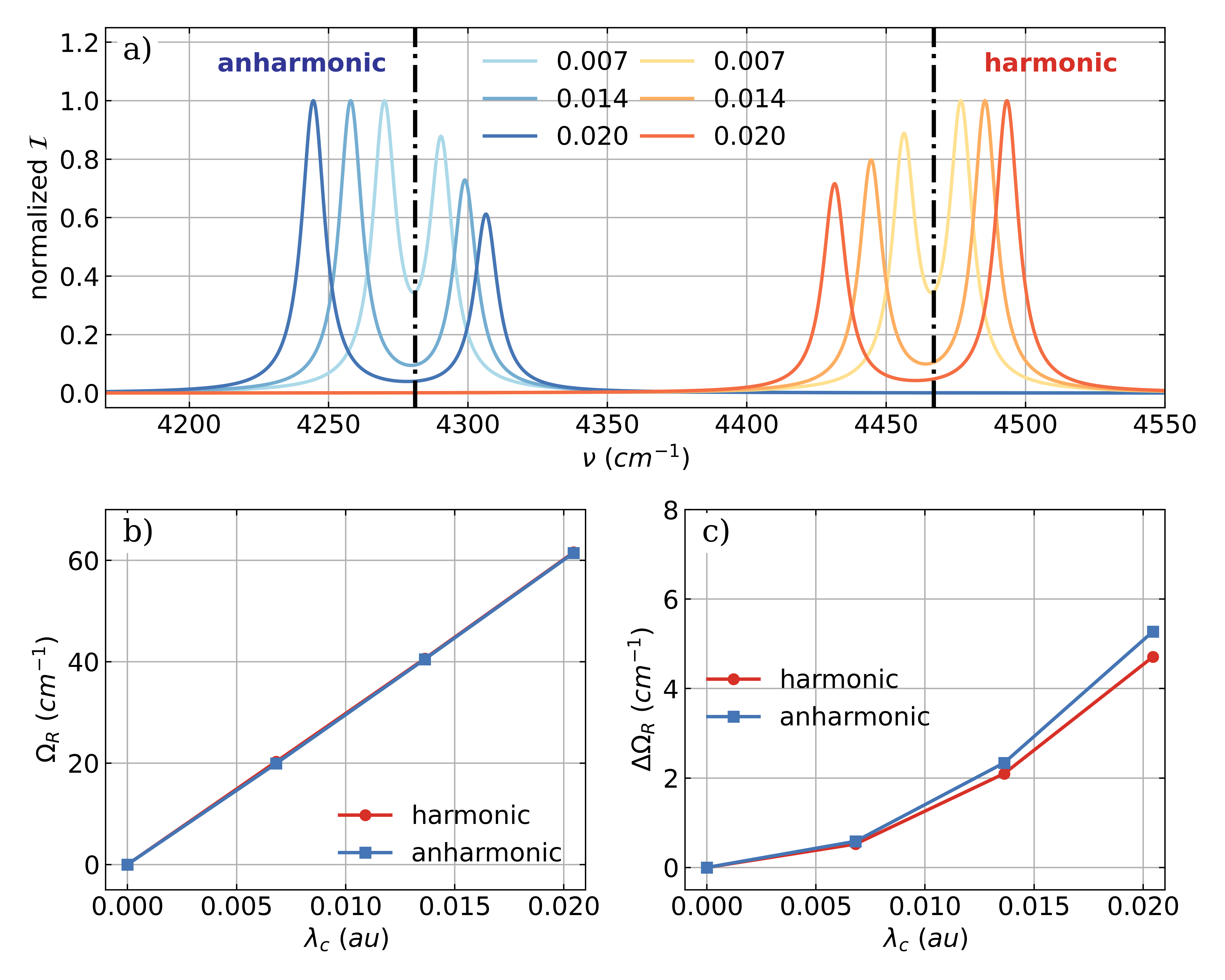}
    \caption{a) Vibro-polaritonic IR spectra of a two antiparallel \ce{HF} molecules calculated in the harmonic approximation (redish) and in the full anharmonic setup (bluish). Black dashed-dotted lines indicate the frequencies of the harmonic (\SI{4467}{\per\centi\meter}) and anharmonic (\SI{4281}{\per\centi\meter}) fundamental transitions. The cavity frequency $\omega_c$ is resonant with the corresponding fundamental transition in both cases, and the coupling strength $\lambda_{c}$ is increased from \SI{0.009}{\au} to \SI{0.039}{\au} (from lightest to darkest color). b) Rabi splitting $\Omega_R$ as a function of $\epsilon_{c}$. c) Asymmetry $\Delta \Omega_R = \omega_c - 0.5\left( \nu^{LP} + \nu^{UP}\right)$ of the Rabi splitting.} 
\label{fig:spectrum_2a_hf_resonant}
\end{figure}

\begin{figure}[htb!]
     \centering
         \includegraphics[width=0.9\textwidth]{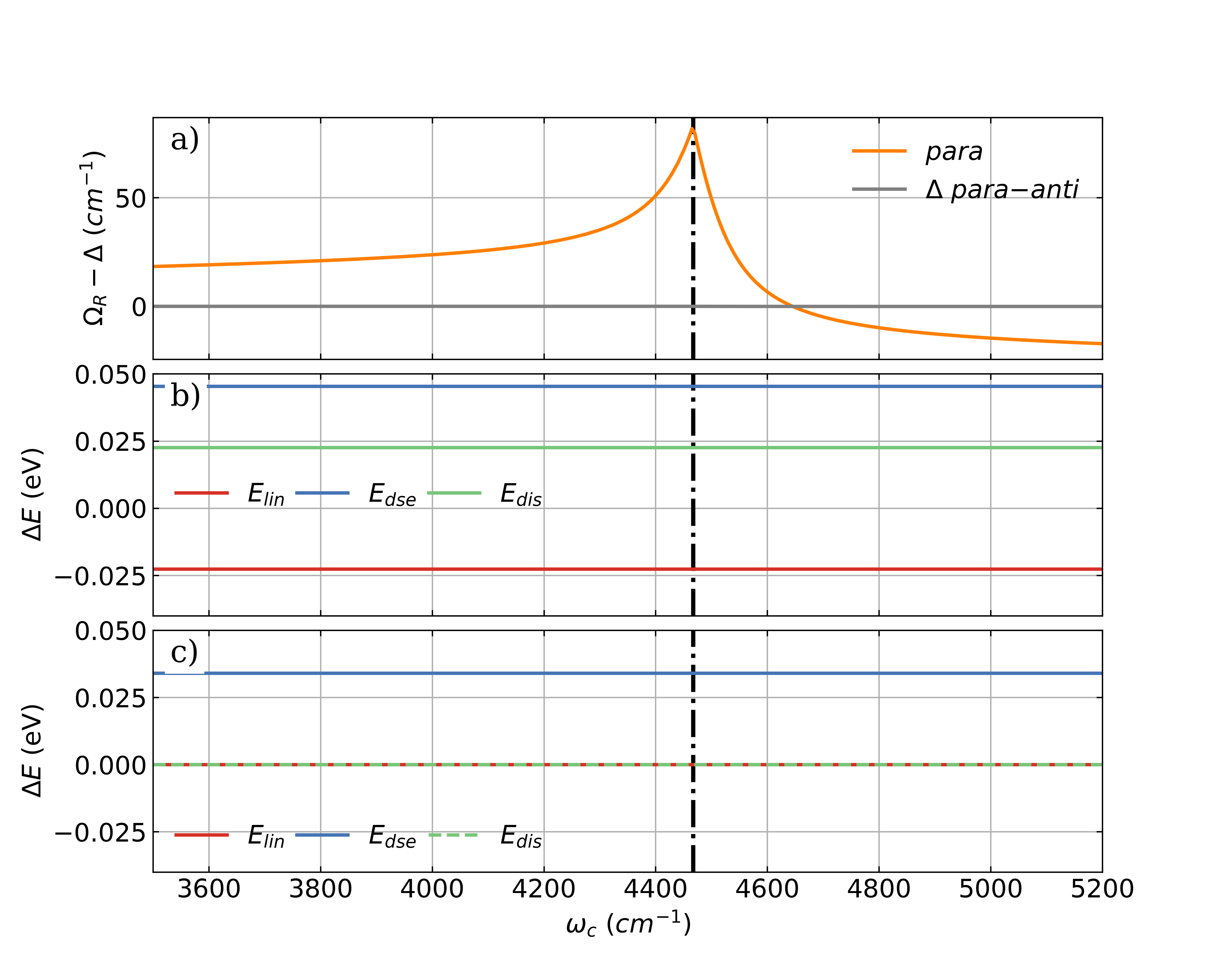}
    \caption{a) Difference between the Rabi splitting $\Omega_R$ and $\Delta =\omega_c - ^{H}\nu_{1}$ as a function of the cavity frequency $\omega_c$ for two parallel aligned \ce{HF} molecules. Here $\Delta$ describes the difference between the cavity frequency $\omega_c$ and the fundamental bare molecular transition  $^{H}\nu_{1}$. The difference of $\Omega_R -\Delta$ between the \textit{all-parallel} and the \textit{antiparallel} configuration is shown in gray. Energy contributions due to the cavity interaction as function of the cavity frequency $\omega_c$ for b) the \textit{all-parallel} orientation and c) the \textit{antiparallel} configuration. Black dashed-dotted lines indicate the frequencies of the harmonic (\SI{4467}{\per\centi\meter}) fundamental transition. A constant coupling strength $\lambda_{c}$ of \SI{0.02}{\au} is used.} \label{fig:detuning_2hf_harm_energy}
\end{figure}

The vibro-polaritonic infrared spectra for different coupling strengths calculated in the harmonic approximation and full-quantum (anharmonic) as well as the trends in the Rabi splitting frequency $\Omega_R$ are shown for the  \textit{all-parallel} configuration of two \ce{HF} molecules in Fig.~\ref{fig:spectrum_2p_hf_resonant} and for the \textit{antiparallel} configuration of two \ce{HF} molecules in Fig.~\ref{fig:spectrum_2a_hf_resonant}. Due to the rescaling of $\bm{\lambda}_{c}$ the observed spectra are nearly identical to the ones obtained for a single \ce{HF} molecule. Consistent with our findings on the properties of ensembles of molecules coupled to an optical cavity~\cite{Schnappinger2023-hh}, the vibro-polaritonic IR spectra are nearly indistinguishable for the two configurations studied. The only noticeable difference is the asymmetry of the Rabi splitting $\Delta \Omega_R = \omega_c - 0.5\left( \nu^{LP} + \nu^{UP}\right)$ shown in Fig.~\ref{fig:spectrum_2p_hf_resonant}~c) and Fig.~\ref{fig:spectrum_2a_hf_resonant}~c). The value of $\Delta \Omega_R$ is slightly smaller for the \textit{antiparallel} configuration. Fig.~\ref{fig:detuning_2hf_harm_energy} shows the difference between $\Omega_R$ and $\Delta =\omega_c - ^{H}\nu_{1}$ in the harmonic approximation as well as the cavity-induced energy changes $E_{lin}$, $E_{dse}$ and $E_{dis}$ as a function of the cavity frequency $\omega_c$, keeping $\lambda_c$ fixed for both configurations.  Also for the bimolecular case, the largest Rabi splitting $\Omega_R$ is obtained for $\omega_c$ resonant with the fundamental molecular transition ($^{H}\nu_{1}$). In agreement with the single molecule results, the difference between $\Omega_R$ and the detuning $\Delta$ is not symmetric with respect to $^{H}\nu_{1}$ and tends to a finite nonzero value even for large detunings, see Fig.~\ref{fig:detuning_2hf_harm_energy}~a). As for the resonant spectra (Fig.~\ref{fig:spectrum_2p_hf_resonant} and Fig.~\ref{fig:spectrum_2a_hf_resonant}), the effect of cavity detuning is the same for both configurations, see gray line Fig.~\ref{fig:detuning_2hf_harm_energy}~a). The cavity-induced energy modifications $E_{lin}$, $E_{dse}$ and $E_{dis}$ shown in Fig.~\ref{fig:detuning_2hf_harm_energy}~b) for the \textit{all-parallel} configuration and in Fig. ~\ref{fig:detuning_2hf_harm_energy}~c) for the \textit{antiparallel} configuration are different for each configuration, but are constant for all values of $\omega_c$. 

\clearpage

\section{Nuclear-Photonic Eigenfunctions vs Vibro-polaritonic Normal Modes}

In this section, we compare the vibro-polaritonic normal mode vectors $\bm{a}_i$ obtained in the harmonic approximation with the nuclear-photonic eigenfunctions $\chi$ calculated full quantum for the case of a single \ce{HF} molecule and for two parallel oriented \ce{HF} molecules.
Before we start the direct comparison, we need to define how to visualize both the normal mode and the eigenfunctions. Since \ce{HF} is a diatomic molecule and the molecular bond is aligned with the $z$ axis of the laboratory frame, we can represent the normal mode vectors $\bm{a}_i$ of size $\left( 3N_{A} +1 \right)$ in a two-dimensional subspace spanned by the $z$ axis of the laboratory frame, which includes both the $z$ components of the \ce{H} atom(s) and the \ce{F} atom(s) and the photon displacement coordinate $q_c$. The nuclear-photonic eigenfunctions are represented as probability densities. For the single-molecule case, the densities are plotted in the full two-dimensional space spanned by the internal (bond length) coordinate $R$ and the photon displacement coordinate $q_c$. To visualize the densities for the two \ce{HF} molecules, all possible two-dimensional projections of the full three-dimensional space spanned by the two internal coordinates ($R_1$ and $R_2$) and $q_c$ are used.

\begin{figure}[htb!]
     \centering
         \includegraphics[width=0.9\textwidth]{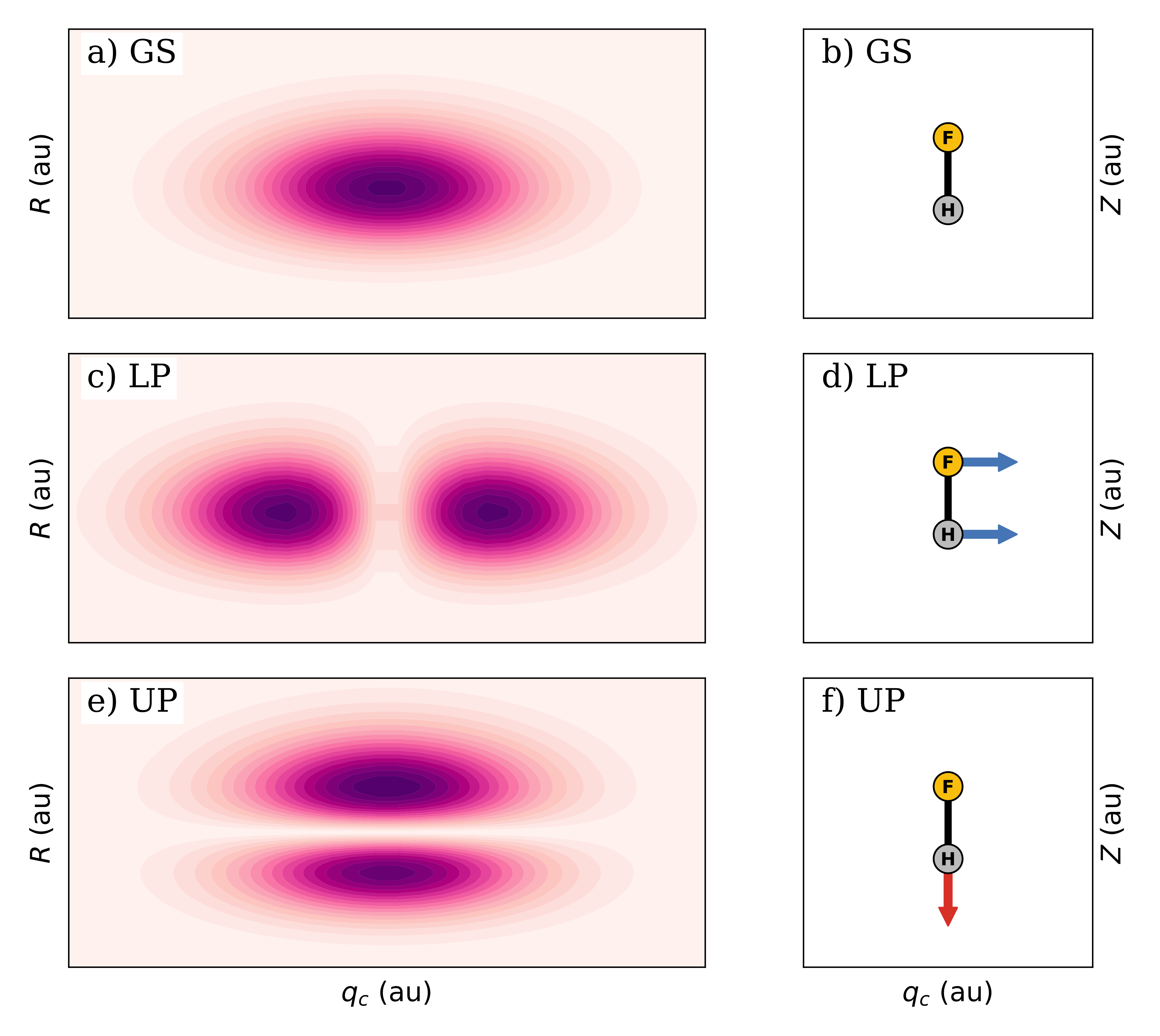}
    \caption{a), c), and d) first three eigenfunctions for a single \ce{HF} molecule described on two-dimensional \gls{cbohf} surface.  b) classic nuclear configuration, d), and e) both mass-weighted normal modes  of a single \ce{HF} molecule in harmonic approximation. The cavity frequency $\omega_c$ is resonant with the corresponding fundamental transition (harmonic \SI{4467}{\per\centi\meter} and anharmonic \SI{4281}{\per\centi\meter}). The coupling strength $\lambda_{c}$ is set to zero.} 
\label{fig:ec0_1p}
\end{figure}

\begin{figure}[htb!]
     \centering
         \includegraphics[width=0.9\textwidth]{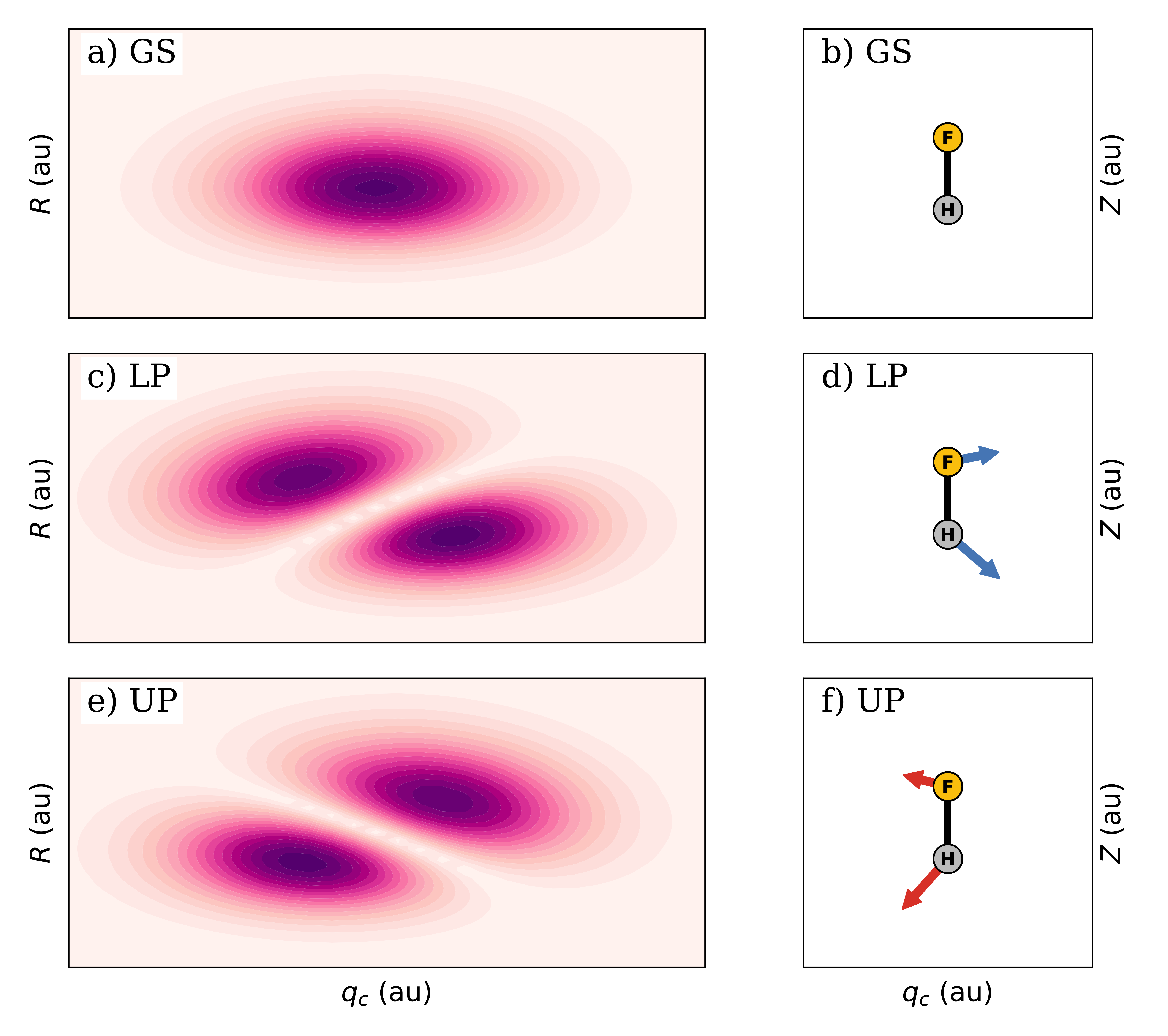}
    \caption{a), c), and d) first three eigenfunctions for a single \ce{HF} molecule described on two-dimensional \gls{cbohf} surface. b) classic nuclear configuration, d), and e) both both mass-weighted normal modes of a single \ce{HF} molecule in harmonic approximation. The cavity frequency $\omega_c$ is resonant with the corresponding fundamental transition (harmonic \SI{4467}{\per\centi\meter} and anharmonic \SI{4281}{\per\centi\meter}). The coupling strength $\lambda_{c}$ is set to \SI{0.019}{\au}.} 
\label{fig:ec3_1p}
\end{figure}

In Fig.~\ref{fig:ec0_1p} both the nuclear-photonic eigenfunctions and the vibro-polaritonic normal modes for a single \ce{HF} molecule and a single cavity mode are shown. The cavity frequency $\omega_c$ is resonant with the corresponding fundamental transition (harmonic \SI{4467}{\per\centi\meter} and anharmonic \SI{4281}{\per\centi\meter}) but the coupling strength $\lambda_{c}$ is set to zero. The to-be \gls{lp} eigenfunction Fig.~\ref{fig:ec0_1p}~c) is a pure one-photon cavity vacuum state with a nodal plan orthogonal to the $q_c$ coordinate and the to-be \gls{up} eigenfunction Fig.~\ref{fig:ec0_1p}~e) is the pure first excited molecular vibrational eigenstate with a nodal plan orthogonal to the $R$ coordinate. The vibro-polaritonic normal modes give the classical pendants. The to-be \gls{lp} normal mode (reduced normal mode shown in Fig.~\ref{fig:ec0_1p}~d)) has only a contribution along the photon displacement coordinate $q_c$, for visualization purposes shown on both atoms. The to-be \gls{up} normal mode (reduced normal mode shown in Fig.~\ref{fig:ec0_1p}~f)) has only a contribution along the $z$ coordinate axis in the laboratory frame, which describes the stretching mode of the \ce{HF} molecule. Since we are discussing mass-weighted normal modes, the vector describing the stretching mode, shown in Fig.~\ref{fig:ec0_1p}~f), is predominantly localized on the hydrogen atom. The eigenfunctions and normal modes in the case of a single molecule for a coupling strength $\lambda_{c}$ of \SI{0.019}{\au} are shown in Fig.~\ref{fig:ec0_1p}. In the full-quantum eigenfunction picture (see Figs~\ref{fig:ec3_1p} c) and e)) the light-matter interaction manifests itself through a rotation of the probability densities in the $q_c$–$R$-plane accompanied by a slight deformation. As discussed in the manuscript (see Eq.~18) and in the literature~\cite{Fischer2021-eq} the eigenfunctions of the \gls{lp} state and the \gls{up} state can be interpreted as linear combinations of the uncoupled one-photon cavity vacuum state and the first excited molecular vibrational state. The mixed characters of the \gls{lp} state and \gls{up} state are also visible in the normal mode representation (see Fig~\ref{fig:ec3_1p} d) and f)). The reduced vectors in the coupled situation are linear combinations of the uncoupled vectors (Fig.~\ref{fig:ec0_1p}~d) and f)). When analyzing not only the direction of the \gls{lp} vector and the \gls{up} vector, but also their length, another effect of the cavity interaction is observed. For both normal modes in the coupled system, the mixing of light and matter leads to the situation that both vectors have significant components on both the \ce{H} and \ce{F} atoms, in contrast to the uncoupled case, see Fig.~\ref{fig:ec0_1p}~f).  

\begin{figure}[htb!]
     \centering
         \includegraphics[width=0.9\textwidth]{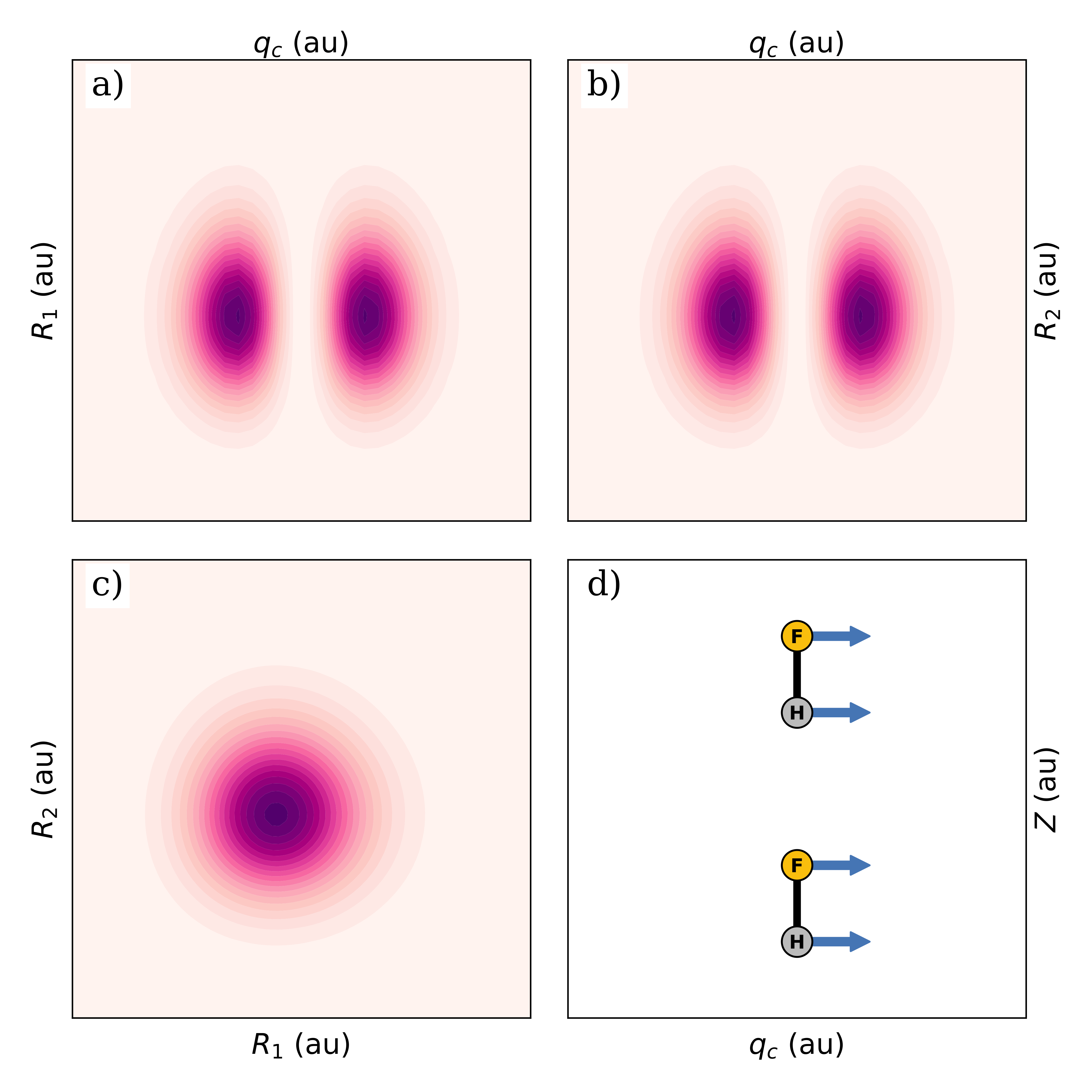}
    \caption{First excited state eigenfunction for  two parallel \ce{HF} molecules described on a three-dimensional \gls{cbohf} surface. For a), b), and c, the eigenfunction is integrated over one coordinate each. d) classic nuclear configuration and first mass-weighted normal mode of two parallel \ce{HF} molecules. The molecules are separated by a distance of \SI{800}{\angstrom}. The cavity frequency $\omega_c$ is resonant with the corresponding fundamental transition (harmonic \SI{4467}{\per\centi\meter} and anharmonic \SI{4281}{\per\centi\meter}). The coupling strength $\lambda_{c}$ is set to zero.} 
\label{fig:ec0_lp_2p}
\end{figure}

\begin{figure}[htb!]
     \centering
         \includegraphics[width=0.9\textwidth]{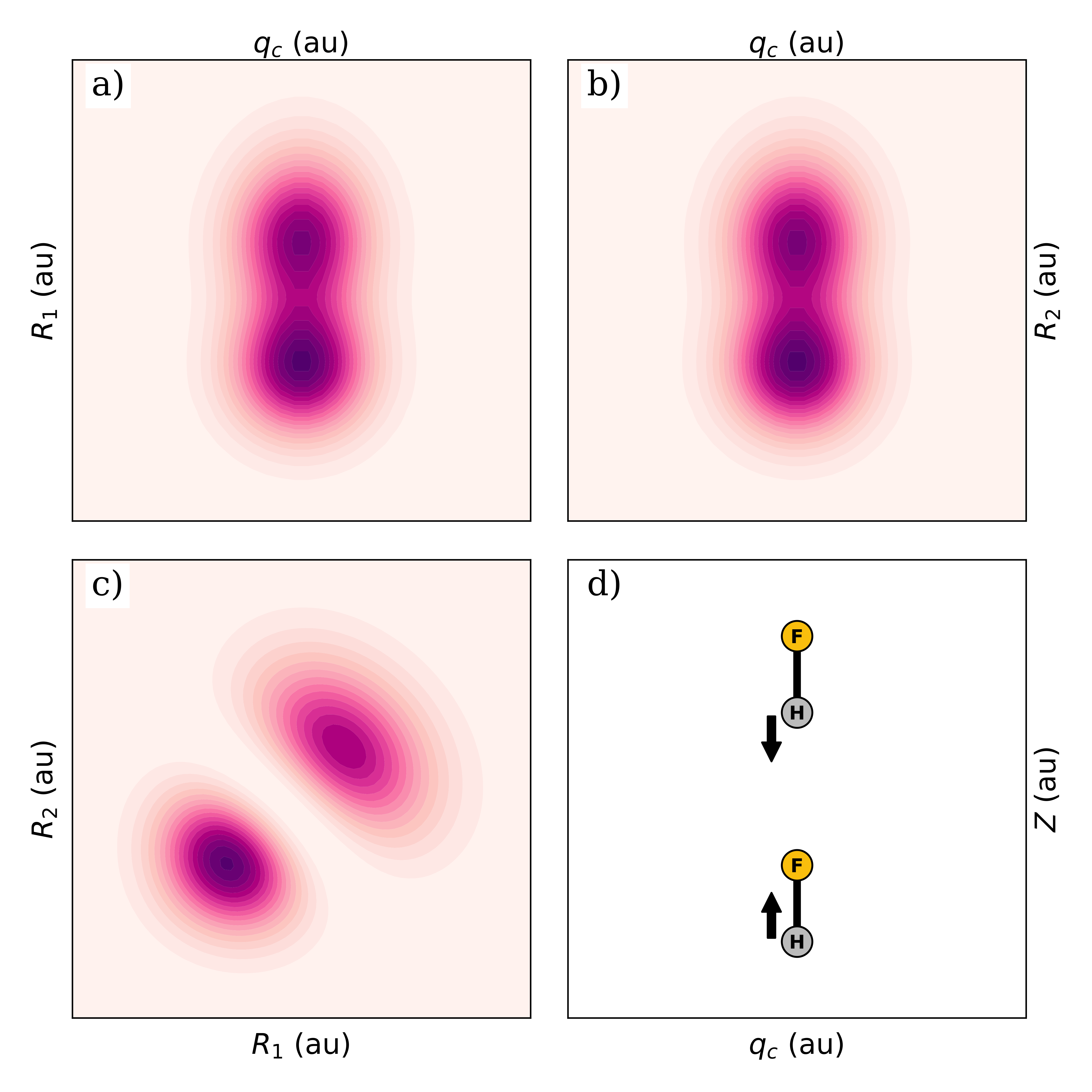}
    \caption{Second excited state eigenfunction for  two parallel \ce{HF} molecules described on a three-dimensional \gls{cbohf} surface. For a), b), and c, the eigenfunction is integrated over one coordinate each. d) classic nuclear configuration and second mass-weighted normal mode of two parallel \ce{HF} molecules. The molecules are separated by a distance of \SI{800}{\angstrom}. The cavity frequency $\omega_c$ is resonant with the corresponding fundamental transition (harmonic \SI{4467}{\per\centi\meter} and anharmonic \SI{4281}{\per\centi\meter}). The coupling strength $\lambda_{c}$ is set to zero.} 
\label{fig:ec0_mp_2p}
\end{figure}

\begin{figure}[htb!]
     \centering
         \includegraphics[width=0.9\textwidth]{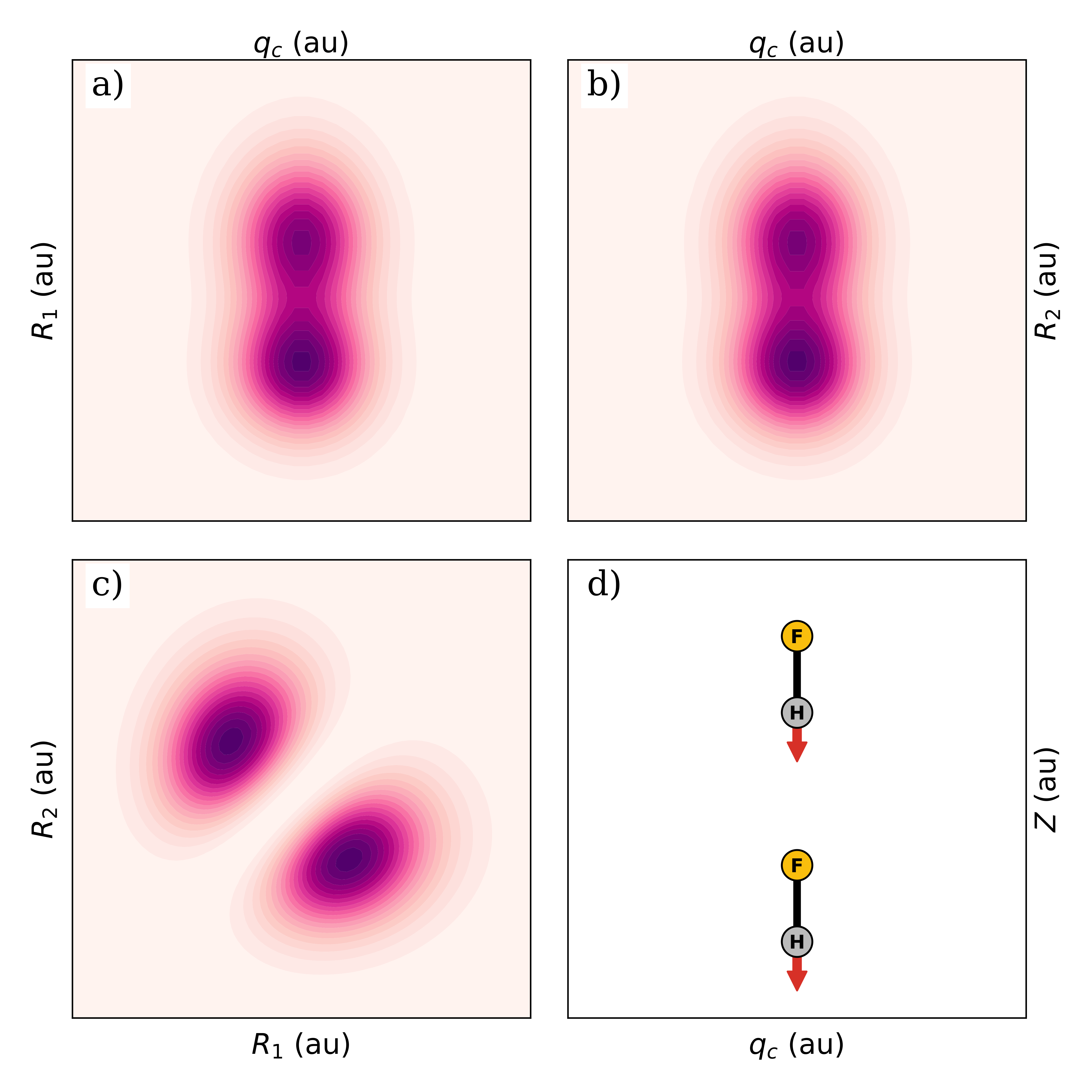}
    \caption{Third excited state eigenfunction for  two parallel \ce{HF} molecules described on a three-dimensional \gls{cbohf} surface. For a), b), and c, the eigenfunction is integrated over one coordinate each. d) classic nuclear configuration and third mass-weighted normal mode of two parallel \ce{HF} molecules. The molecules are separated by a distance of \SI{800}{\angstrom}. The cavity frequency $\omega_c$ is resonant with the corresponding fundamental transition (harmonic \SI{4467}{\per\centi\meter} and anharmonic \SI{4281}{\per\centi\meter}). The coupling strength $\lambda_{c}$ is set to zero.} 
\label{fig:ec0_up_2p}
\end{figure}

\begin{figure}[htb!]
     \centering
         \includegraphics[width=0.9\textwidth]{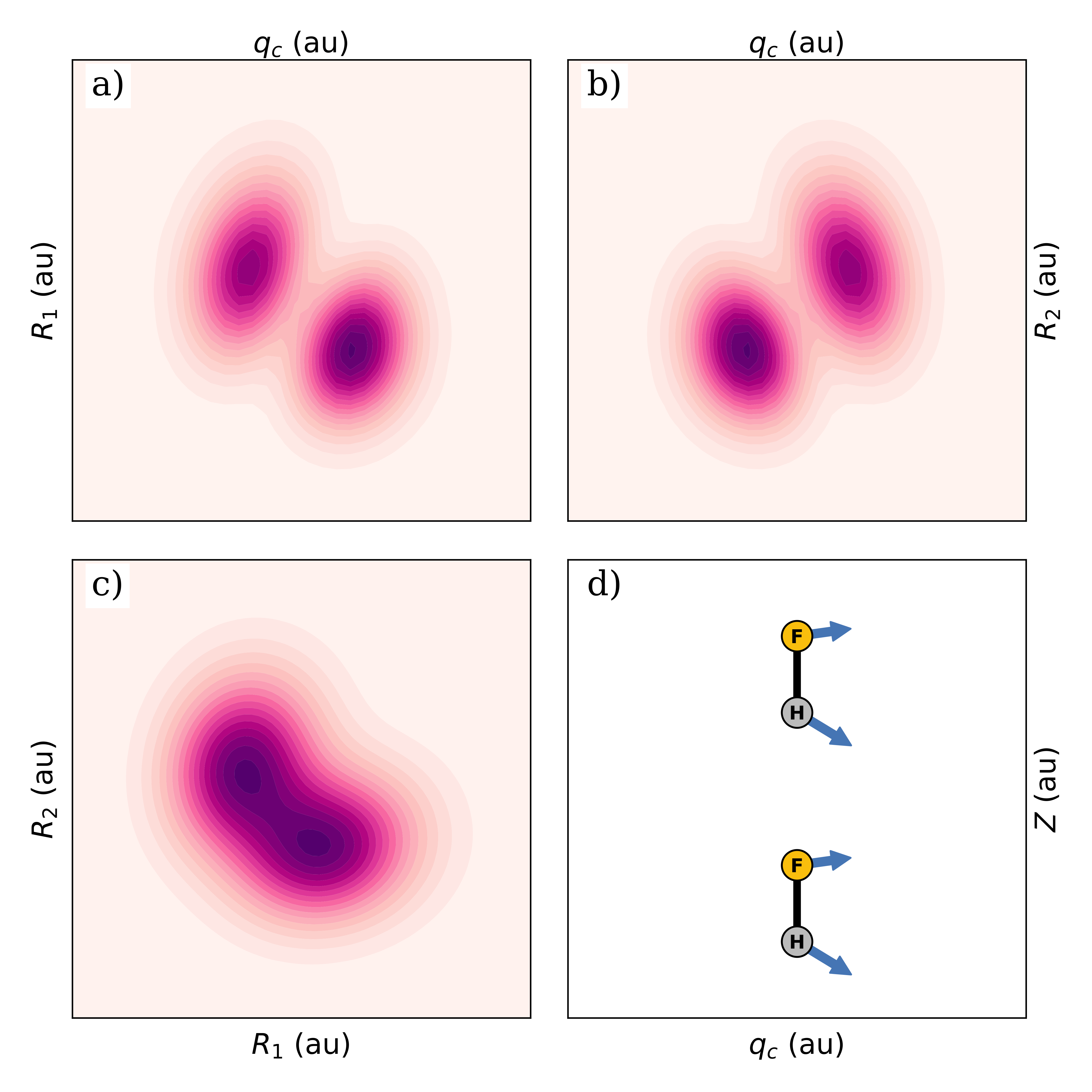}
    \caption{First excited state eigenfunction for  two parallel \ce{HF} molecules described on a three-dimensional \gls{cbohf} surface. For a), b), and c, the eigenfunction is integrated over one coordinate each. d) classic nuclear configuration and first mass-weighted normal mode of two parallel \ce{HF} molecules. The molecules are separated by a distance of \SI{800}{\angstrom}. The cavity frequency $\omega_c$ is resonant with the corresponding fundamental transition (harmonic \SI{4467}{\per\centi\meter} and anharmonic \SI{4281}{\per\centi\meter}). The coupling strength $\lambda_{c}$ is set to \SI{0.019}{\au}.} 
\label{fig:ec3_lp_2p}
\end{figure}

\begin{figure}[htb!]
     \centering
         \includegraphics[width=0.9\textwidth]{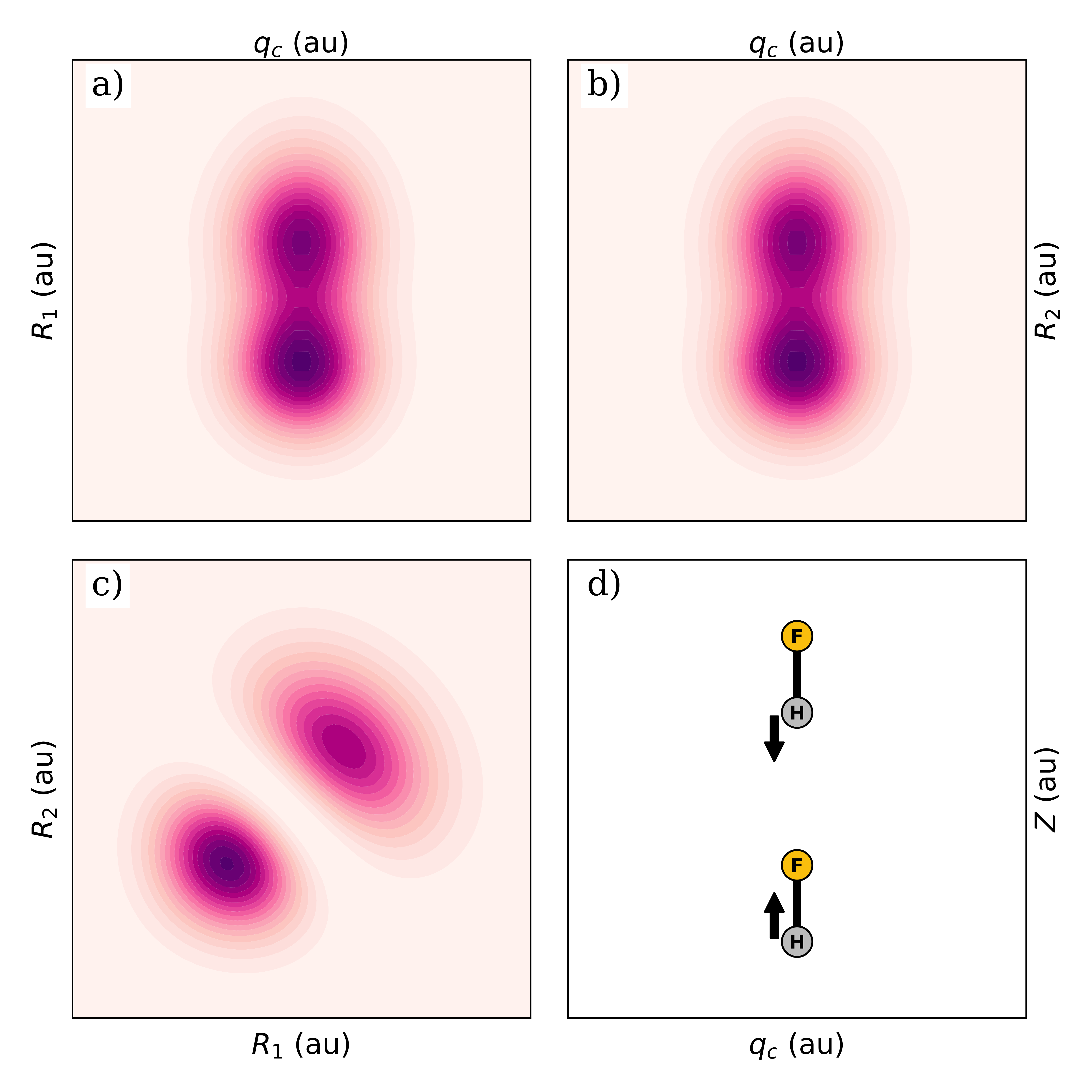}
    \caption{Second excited state eigenfunction for  two parallel \ce{HF} molecules described on a three-dimensional \gls{cbohf} surface. For a), b), and c, the eigenfunction is integrated over one coordinate each. d) classic nuclear configuration and second mass-weighted normal mode of two parallel \ce{HF} molecules. The molecules are separated by a distance of \SI{800}{\angstrom}. The cavity frequency $\omega_c$ is resonant with the corresponding fundamental transition (harmonic \SI{4467}{\per\centi\meter} and anharmonic \SI{4281}{\per\centi\meter}). The coupling strength $\lambda_{c}$ is set to \SI{0.019}{\au}.} 
\label{fig:ec3_mp_2p}
\end{figure}

\begin{figure}[htb!]
     \centering
         \includegraphics[width=0.9\textwidth]{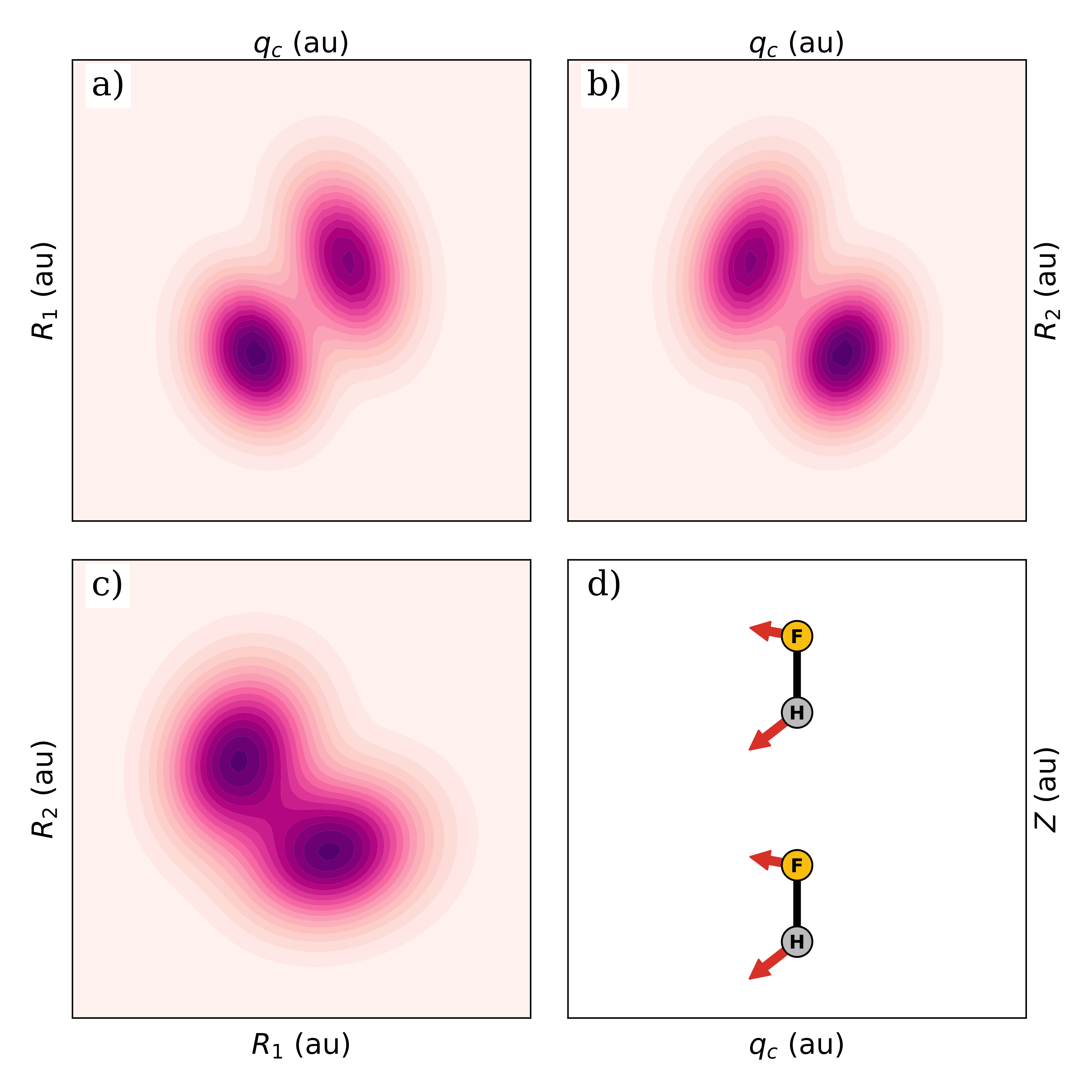}
    \caption{Third excited state eigenfunction for  two parallel \ce{HF} molecules described on a three-dimensional \gls{cbohf} surface. For a), b), and c, the eigenfunction is integrated over one coordinate each. d) classic nuclear configuration and third mass-weighted normal mode of two parallel \ce{HF} molecules. The molecules are separated by a distance of \SI{800}{\angstrom}. The cavity frequency $\omega_c$ is resonant with the corresponding fundamental transition (harmonic \SI{4467}{\per\centi\meter} and anharmonic \SI{4281}{\per\centi\meter}). The coupling strength $\lambda_{c}$ is set to \SI{0.019}{\au}.} 
\label{fig:ec3_up_2p}
\end{figure}

For the two parallel aligned \ce{HF} molecules nuclear-photonic eigenfunctions and vibro-polaritonic normal modes for the \gls{lp} state, the dark middle-polariton state and the \gls{up} state are shown in Fig.~\ref{fig:ec0_lp_2p}, Fig. ~\ref{fig:ec0_mp_2p}, Fig.~\ref{fig:ec0_up_2p} for the uncoupled system ( $\lambda_{c} = 0.0$) and in Fig. ~\ref{fig:ec3_lp_2p}, Fig.~\ref{fig:ec3_mp_2p}, Fig.~\ref{fig:ec3_up_2p} for a coupling strength $\lambda_{c}$ of \SI{0.019}{\au}. For the \gls{lp} and \gls{up} states, identical behavior is observed when coupled to a cavity compared to the case of a single \ce{HF}. In the uncoupled system, the second vibrational transition is dark and remains dark when coupled to a cavity. Therefore, the underlying eigenfunctions and normal modes are unchanged; see (Fig.~\ref{fig:ec0_mp_2p}) and (Fig.~\ref{fig:ec3_mp_2p}).  

Overall, both descriptions, the nuclear-photonic eigenfunctions and the vibro-polaritonic normal modes, contain similar information about the formation of hybrid-light matter states.

\clearpage

\section{Beyond Diatomic Molecules: Vibro-Polaritonic Spectra of \ce{NH3}}

\begin{figure}[htb!]
     \centering
         \includegraphics[width=0.9\textwidth]{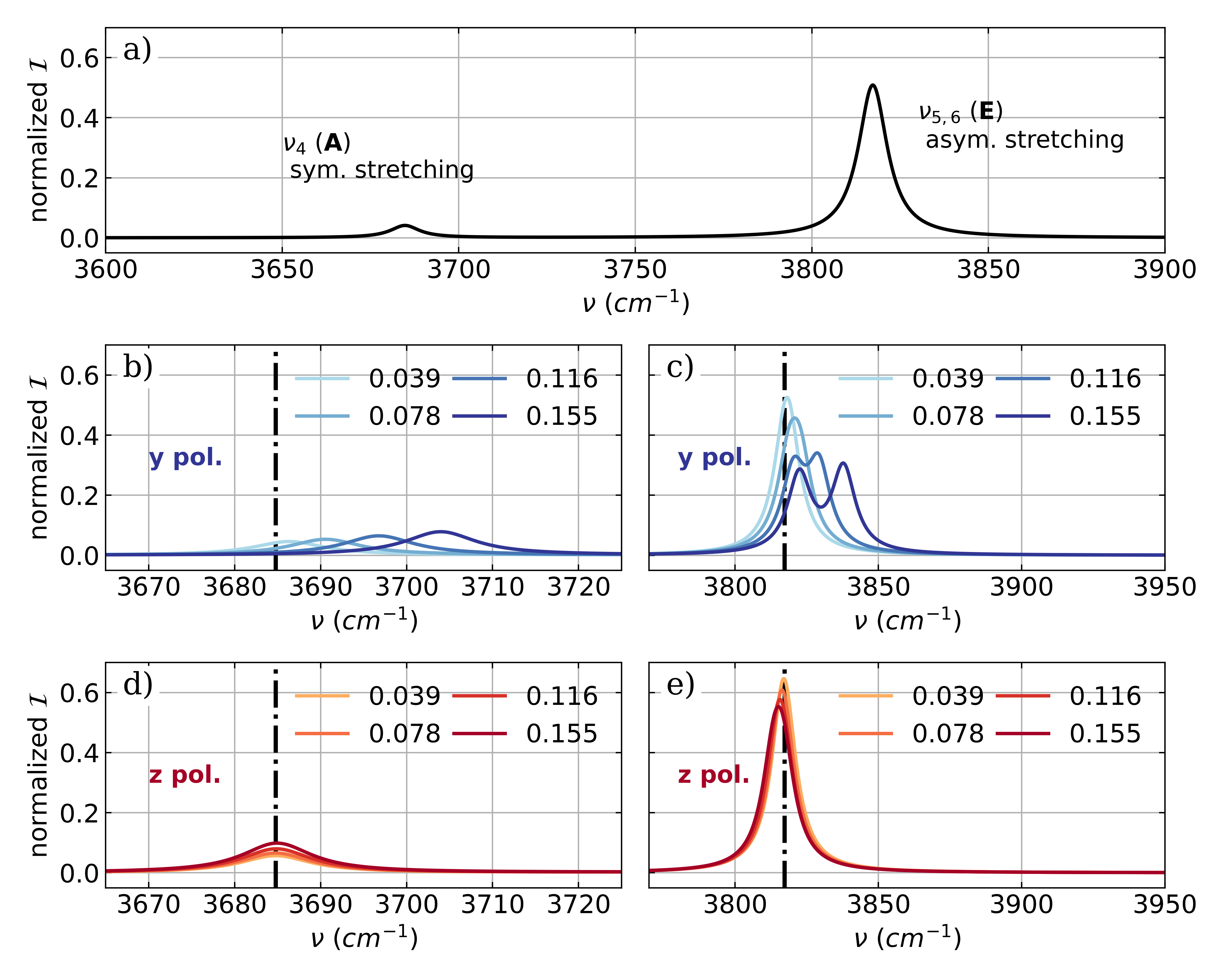}
    \caption{a) High energy part of the vibronic IR spectra of a single \ce{NH3} molecule calculated in the harmonic approximation. High energy part of the vibro-polaritonic IR spectra of a single \ce{NH3} molecule zoomed into the symmetric stretching (b) and d)) and asymmetric stretching modes (c) and e)). The polarization axis of the cavity mode is the $y$ axis for b) and c) and the $z$ axis for d) and e). The cavity frequency $\omega_c$ is resonant with the symmetric bending mode  (\SI{1103}{\per\centi\meter}) and the cavity field strength $\lambda_{c}$ is increased from \SI{0.039}{\au} to \SI{0.155}{\au}.}  
\label{fig:spectrum_nh3_v1_stretch}
\end{figure}

The bare molecular vibronic IR spectrum and the vibro-polaritonic IR spectra of a
single \ce{NH3} molecule are shown in Fig.~\ref{fig:spectrum_nh3_v1_stretch} for the three stretching modes. The cavity frequency $\omega_c$ is set to be resonant with the symmetric bending mode $\nu_1$ (\SI{1103}{\per\centi\meter}). The vibro-polaritonic IR spectra are calculated for the $y$ polarization (Fig.~\ref{fig:spectrum_nh3_v1_stretch}~b) and d)) and the $z$ polarization (Fig.~\ref{fig:spectrum_nh3_v1_stretch}~c) and d)) of the single cavity mode. Since the chosen cavity frequency $w_c$ is off-resonant with the stretching modes, the cavity interaction leads to only modifications of the signals by changing the energetics and breaking the molecular symmetries. These effects are stronger for the $y$ polarization because of the molecular orientation chosen. 

\clearpage

\bibliography{lit.bib}
\providecommand{\latin}[1]{#1}
\makeatletter
\providecommand{\doi}
  {\begingroup\let\do\@makeother\dospecials
  \catcode`\{=1 \catcode`\}=2 \doi@aux}
\providecommand{\doi@aux}[1]{\endgroup\texttt{#1}}
\makeatother
\providecommand*\mcitethebibliography{\thebibliography}
\csname @ifundefined\endcsname{endmcitethebibliography}
  {\let\endmcitethebibliography\endthebibliography}{}